\documentclass[aps,showpacs]{revtex4}
\usepackage{graphicx}
\usepackage[all]{xy}
\usepackage{amsmath}
\usepackage{amssymb}
\usepackage{color}
\def\ssc{\scriptscriptstyle}
\newcommand{\be}{\begin{equation}}
\newcommand{\ee}{\end{equation}}
\newcommand{\ben}{\begin{eqnarray}}
\newcommand{\een}{\end{eqnarray}}
\newcommand{\bes}{\begin{subequations}}
\newcommand{\ees}{\end{subequations}}
\newcommand{\wt}{\widetilde}

\newcommand{\bb}{\bibitem}
\begin{document}
\title{A simple and direct method for generating travelling\\ wave solutions for nonlinear equations}
\author{D. Bazeia,$^a$ Ashok Das,$^{b,c}$ L. Losano,$^a$ and A. Silva$^{a,d}$}
\affiliation{{\small {{$^a$}{Departamento de F\'{\i}sica, Universidade Federal da Para\'{\i}ba, 58051-970 Jo\~ao Pessoa PB, Brazil}}\\
{$^b$}{Department of Physics and Astronomy, University of Rochester, Rochester, NY 14627-0171} \\
{$^c$}{Saha Institute of Nuclear Physics, 1/AF Bidhannagar, Calcutta 700064, India}\\
{$^d$}{Departamento de F\'\i sica, Universidade Estadual da Para\'\i ba, Campina Grande PB, Brazil}}}
\begin{abstract}
We propose a simple and direct method for generating travelling wave solutions for nonlinear integrable equations. We illustrate how nontrivial solutions for the KdV, the mKdV and the Boussinesq equations can be obtained from simple solutions of linear equations. We describe how using this method, a soliton solution of the KdV equation can yield soliton solutions for the mKdV as well as the Boussinesq equations. Similarly, starting with cnoidal solutions of the KdV equation, we can obtain the corresponding solutions for the mKdV as well as the Boussinesq equations. Simple solutions of linear equations can also lead to cnoidal solutions of nonlinear systems. Finally, we propose and solve some new families of KdV equations and show how soliton solutions are also obtained for the higher order equations of the KdV hierarchy using this method.
\end{abstract}


\maketitle

\section{Introduction}
\label{s1}

Nonlinear integrable systems have been studied vigorously from a variety of points of view. Since the KdV equation was proposed \cite{kdv1}, the subject of integrable models has developed in many directions - the Lax representation, inverse scattering method, B\"{a}cklund transformations, Zakharov-Shabat formulation, quantum integrability and quantum groups and so on \cite{b1,b2,b3,b4,b5,b6,b7}. Each of these developments is significant in its own right and has opened up new areas of research -- see, e.g., \cite{kdv1,b1,b2,b3,b4,b5,b6,b7,mkdv,lax,faddeev1,hirota,zs,zs1,gardner,ablo,ablo1,lax1,mat}. The solution of a nonlinear integrable equation is normally obtained through the inverse scattering method \cite{lax,faddeev1,ablo1}. In some sense, this can be thought of as the ``Fourier transform" method for nonlinear systems. Here one tries to associate  with a given nonlinear integrable equation a pair of linear equations (with the dynamical variable of the nonlinear system as the potential) for which the ``time" evolution of various quantities of interest is much simpler. By studying the ``time" evolution of the scattering data in this linear system, one reconstructs the time evolution of the dynamical variable (the potential) through the method of inverse scattering. This is a powerful, indirect method for solving nonlinear integrable systems. There also exists a more direct way of solving a nonlinear integrable system that goes under the name of Hirota's bilinear method \cite{hirota}. Here one introduces a non-conventional derivative (a derivative that acts forwards and backwards with a relative sign which is quite common in the study of quantum field theory) and rewrites the nonlinear equation as a linear equation in two variables which then leads to a direct solution of the original system. While both these methods are very powerful, they are quite technical.

In this paper, we would like to propose an alternative direct method for solving nonlinear integrable systems that is quite simple. In this method, one derives a nontrivial solution of a nonlinear integrable system from a known solution of a linear equation or another nonlinear equation. In the generic sense, therefore, this method can be thought of as a ``B\"{a}cklund" transformation method. However, unlike the conventional B\"{a}cklund transformation method where one has to deal with zero curvature conditions, here the construction of the map is much more straight forward and, therefore, we believe that it is quite useful. This transformation method was already studied in connection with relativistic scalar field theories to generate nontrivial classical solutions to complicated theories from known ones in a simpler theory \cite{dd}. In that context, it was also termed the ``deformation" method, and in \cite{dd1} it was shown to work nicely for a variety of situations. However, the power of the method has led us to propose it as a useful direct method for obtaining solutions to nonlinear integrable systems.

The paper is organized as follows. In section {\bf II}, we illustrate the method within the context of a general Hamiltonian equation which is third order in space derivatives. In section {\bf III} we show examples of how nontrivial travelling solutions for the KdV, the mKdV \cite{mkdv} as well as the Boussinesq \cite{bou} equations can be obtained from simple solutions of a third order linear equation. In section {\bf IV}, soliton solutions for the mKdV as well as the Boussinesq equations are obtained from the soliton solution of the KdV equation. In section {\bf V}, cnoidal solutions for the mKdV as well as the Boussinesq equations are derived from that of the KdV equation. We indicate briefly in this section how cnoidal solutions of nonlinear equations can be obtained from simple solutions of linear equations. The method proposed is then used to investigate new equations and the corresponding travelling wave solutions in section {\bf VI}, and we conclude this work with some open questions in section {\bf VII}.

\section{Essentials of the method}

Let us consider a general Hamiltonian equation in $1+1$ dimensions of third order in the space derivatives of the form
\begin{equation}
u_{t} + \left(f(u)\right)_{x} + \alpha u_{xxx} = 0,\label{general}
\end{equation}
where the subscripts denote derivatives with respect to the particular variables and $\alpha$ is a constant. Here $f(u)$ is a monomial in the dynamical variable $u$ and the equation is linear/nonlinear depending on the nature of this monomial function. Let us assume that this equation possesses a travelling solution of the form
\begin{equation}
u_{\rm\ssc T}  = u (kx-\omega t),\label{travelling}
\end{equation}
where, in general, $\omega = \omega (k)$ with the explicit form for the dispersion determined from the structure of the equation and in the case of a nonlinear equation, even the amplitude of the solution (height of the wave) is a function of $k$ and, therefore, of $\omega$. In the case of the KdV equation with $f(u) = 3u^{2}, \alpha=1$ (the coefficients in the KdV equation can be rescaled to any value through appropriate transformations)
\begin{equation}
u_{t} + 6 u u_{x} + u_{xxx} = 0,\label{kdv}
\end{equation} 
for example, the one soliton travelling solution has the form
\begin{equation}\label{form4}
u_{\rm\ssc T}  = A\ {\rm sech}^{2}\ (kx - \omega t),\quad \omega= 4k^{3},\quad A = 2k^{2} =\left(\frac{\omega^{2}}{2}\right)^{\frac{1}{3}}.
\end{equation}
This can also be rewritten in the more familiar form
\begin{equation}
u_{\rm\ssc T} = 2k^{2}\ {\rm sech}^{2}\ k (x-4k^{2}t) = \frac{c}{2}\ {\rm sech}^{2}\ k (x - ct),\label{soliton}
\end{equation}
where we have identified the speed of the travelling solution as $c = 4k^{2}$. We note here that if we transform $u\rightarrow -u$, the KdV equation takes the form
\begin{equation}
u_{t} - 6uu_{x} + u_{xxx} =0,
\end{equation}
for which the soliton solution \eqref{soliton} has the inverted form
\begin{equation}
u_{\rm\ssc T} = - 2k^{2}\ {\rm sech}^{2}\ k (x-4k^{2}t).
\end{equation}

A travelling solution of \eqref{general} of the form \eqref{travelling} then satisfies the equation
\begin{equation}
-\frac{\omega}{k} u_{x} + \left(f(u)\right)_{x} + \alpha u_{xxx} = 0,
\end{equation}
which can be integrated to give
\begin{equation}
u_{xx} = \frac{\omega}{\alpha k}\ u - \frac{1}{\alpha}\ f(u) + \beta = V' (u),\label{integrated}
\end{equation}
where $\beta$ is a constant of integration whose value can be determined from the asymptotic (boundary) conditions satisfied by the solution and we have introduced the notation
\be\label{further}
V' (u) = \frac{\partial V (u)}{\partial u}.
\ee
Here a constant of integration may be included in the definition of $V(u),$ which in field theory is called a potential \cite{dd}.

Let us next suppose that we have another dynamical system of third order with the dynamical variable $v$ in $1+1$ dimensions and that we are interested in determining a travelling wave solution of the form
\begin{equation}
v_{\rm\ssc T} = v (kx - \wt\omega t),\label{travelling1}
\end{equation}
where, in general, $\wt\omega = \wt\omega (k)$ will be a different function of $k$. As in the earlier case (see \eqref{integrated} and \eqref{further}),  the travelling solution \eqref{travelling1} brings the dynamical equation for the system to the form
\begin{equation}
v_{xx} = \widetilde{V}' (v) = \frac{\partial \widetilde{V} (v)}{\partial v}.\label{integrated1}
\end{equation}

We now make the important observation that if there is an invertible map
\begin{equation}
u = g (v),\label{map}
\end{equation}
such that \cite{dd}
\begin{equation}
\widetilde{V} (v) = \frac{V (g(v))}{\left(g' (v)\right)^{2}},\quad g' (v) = \frac{\partial g (v)}{\partial v},\label{map1a}
\end{equation}
then, Eq.~(\ref{integrated}) can be compared with \eqref{integrated1} to show that if a travelling solution for the first dynamical system is known, we can obtain a travelling wave solution for the second system simply as
\begin{equation}
v_{\rm\ssc T} = v (kx - \wt\omega t) = g^{-1} (u (kx - \wt\omega t)).\label{solution}
\end{equation}
It is important to note here that the relation \eqref{map} is not required to map the first dynamical equation into the second (whose solution we are interested in constructing). Rather, it relates the derived potentials of the two systems. In this sense, it is different from a ``Miura" map.

We will show, through many examples, that this is a very powerful and useful method for constructing travelling wave solutions for nonlinear systems. As our discussion shows, the dynamical systems need not be integrable for the method to work. However, the utility of the method is clearly  more significant for obtaining travelling solutions for integrable systems. In the next few sections, we will discuss in detail how such solutions are derived in a variety of cases.

\section{Solutions of nonlinear equations from linear systems}

In this section, we illustrate how this procedure works in generating nontrivial solutions for nonlinear equations starting with those for linear systems. Let us assume that $f(u) = 0, \alpha = 1$ so that \eqref{general} takes the form
\begin{equation}
u_{t} + u_{xxx} = 0,\label{linear}
\end{equation}
which is a third order linear equation and has a simple travelling solution of the form
\begin{equation}
u_{\rm\ssc T} = A \cos (kx + \omega t),\quad \omega = k^{3},\label{solution2}
\end{equation}
and the amplitude $A$ is an arbitrary constant. In this case, \eqref{integrated} yields
\begin{equation}
u_{xx} = - \frac{\omega}{k}\ u + \beta = - k^{2}\ u + \beta = V' (u).\label{integrated2}
\end{equation}
Comparing with the solution \eqref{solution2}, it follows that the constant of integration vanishes, $\beta = 0$, so that we have
\begin{equation}
V (u) = - \frac{k^{2}}{2}\ u^{2} + \gamma,\label{potential}
\end{equation}
where $\gamma$ is the constant of integration in the potential.

If we are trying to obtain a nontrivial solution of say the KdV equation \eqref{kdv}
\begin{equation}
v_{t} + 6vv_{x} + v_{xxx} = 0,
\end{equation}
of the form
\begin{equation}
v_{\rm\ssc T} = v (kx - \wt\omega t),
\end{equation}
we can follow the discussion in \eqref{integrated}-\eqref{further} to construct
\begin{equation}
\widetilde{V} (v) = \frac{\wt\omega}{2k}\ v^{2} - v^{3} + \widetilde{\beta} v + \widetilde{\gamma},\label{kdvpotential}
\end{equation}
where $\widetilde{\gamma}$ is the integration constant in the potential.

Next, let us note that if we choose the constant of integration  $\gamma= {A^2k^{2}}/{2}$ for simplicity, then under the map
\begin{equation}
u (kx - \wt\omega t) = g (v (kx - \wt\omega t)) = A \cos\left({\rm sech}^{-1}\left(\sqrt{\frac{v}{2k^{2}}}\right)\right),\label{map1}
\end{equation}
we obtain
\begin{equation}
\frac{V(g(v))}{\left(g'(v)\right)^{2}} = 2k^{2}v^{2} - v^{3}.
\end{equation}
This can be compared with \eqref{kdvpotential} and we note that we can identify ${\widetilde V}(v)$ as in (\ref{map1a}) provided
\begin{equation}
\widetilde{\beta} = \widetilde{\gamma} = 0,\quad \wt\omega = 4k^{3}.\label{constants}
\end{equation}
In this case, the solution for the KdV equation can be obtained as (see \eqref{solution})
\begin{equation}
v_{\rm\ssc T} = g^{-1} (u (kx - \wt\omega t)) = 2k^{2}\ {\rm sech}^{2}\ k(x-4k^{2}t),\label{kdvsolution}
\end{equation}
which we recognize as the one soliton solution of the KdV equation \eqref{soliton} and we note here that the vanishing of the constants of integration in \eqref{constants} is a reflection of the fact that the soliton solution and its derivative vanish asymptotically. In this case, we have constructed this nontrivial solution of the KdV equation from a simple trigonometric solution of the linear equation \eqref{linear}.

On the other hand, if we want to construct a solution of the mKdV equation with the negative sign
\begin{equation}
v_{t} - 6 v^{2}v_{x} +v_{xxx} = 0,\label{mkdv-}
\end{equation}
of the form
\begin{equation}
v_{\rm\ssc T} = v (kx - \wt\omega t),
\end{equation}
then, we can proceed as in \eqref{integrated}-\eqref{further} and determine
\begin{equation}
\widetilde{V} (v) = \frac{\wt\omega}{2k}\ v^{2} + \frac{1}{2}\ v^{4} + \widetilde{\beta} v + \widetilde{\gamma}.\label{mkdv-potential}
\end{equation}
Here $\widetilde{\beta},\widetilde{\gamma}$ are the two constants of integration. It is clear that if we choose as before $\gamma =A^2k^2/2$, then under the map
\begin{equation}
u (kx - \wt\omega t) = g (v (kx-\wt\omega t)) = \pm A \cos \left(\tanh^{-1}\left(\frac{v}{k}\right)\right),\label{map2}
\end{equation}
we obtain
\begin{equation}
\frac{V (g(v))}{\left(g' (v)\right)^{2}} = \frac{k^{4}}{2} - k^{2} v^{2} + \frac{1}{2}\ v^{4},
\end{equation}
which can be compared with \eqref{mkdv-potential}. We note that we can identify ${\widetilde V}(v)$ as in (\ref{map1a})
provided
\begin{equation}
\widetilde{\beta} = 0,\quad \widetilde{\gamma} = \frac{k^{4}}{2},\quad \wt\omega = -2k^{3}.
\end{equation}
In this case, the solution of the negative mKdV equation \eqref{mkdv-} is obtained to be
\begin{equation}
v_{\rm\ssc T} = g^{-1} (u (kx - \wt\omega t)) = \pm k \tanh k(x+2k^{2}t).\label{mkdv-solution}
\end{equation}
This is the well known soliton solution of the negative mKdV that is not localized, but here we have derived it from a simple trigonometric solution of a linear equation. (Parenthetically we note that a nontrivial constant of integration $\widetilde{\gamma}$ in this case signals that the solution does not vanish asymptotically (since it is not localized), but the derivative does.)

The travelling solution for the (positive) mKdV equation
\begin{equation}
v_{t} + 6v^{2}v_{x} + v_{xxx} = 0,\label{mkdv+}
\end{equation}
can also be constructed in a similar manner as follows. For a travelling solution of this equation of the form
\begin{equation}
v_{\rm\ssc T} = v (kx - \wt\omega t),
\end{equation}
following the steps \eqref{integrated}-\eqref{further}, we obtain from \eqref{mkdv+}
\begin{equation}
\widetilde{V} (v) = \frac{\wt\omega}{2k}\ v^{2} - \frac{1}{2}\ v^{4} + \widetilde{\beta} v + \widetilde{\gamma}.\label{mkdv+potential}
\end{equation}
Here $\widetilde{\beta}$ and $\widetilde{\gamma}$ are the constants of integration and this potential can be compared with that in \eqref{mkdv-potential}. We note that, as in the earlier cases if we choose $\gamma = {A^2k^{2}}/{2}$, then under the map
\begin{equation}
u (kx - \wt\omega t) = g (v (kx - \wt\omega t)) = \pm A \cos\left({\rm sech}^{-1}\left(\frac{v}{k}\right)\right),\label{map3}
\end{equation}
we obtain
\begin{equation}
\frac{V (g(v))}{\left(g' (v)\right)^{2}} = \frac{k^{2}v^{2}}{2} - \frac{1}{2}\ v^{4},
\end{equation}
which can be compared with \eqref{mkdv+potential}. It is clear that we can identify ${\widetilde V}(v)$ as in (\ref{map1a}) provided
\begin{equation}
\widetilde{\beta} = \widetilde{\gamma} = 0,\quad \wt\omega = k^{3}.
\end{equation}
In this case, we can determine the travelling solution for the mKdV equation to be
\begin{equation}
v_{\rm\ssc T} = g^{-1} (u (kx - \wt\omega t)) = \pm k\ {\rm sech}\ k(x-k^{2}t),\label{mkdv+solution}
\end{equation} 
which represents the one soliton solution of the mKdV equation. We note here that the two mKdV equations, \eqref{mkdv-} and \eqref{mkdv+}, are invariant under $v\rightarrow -v$, which is the reason for the solutions with both signs in these two cases.

Finally, let us note that we can also construct a travelling solution for the Boussinesq equation in a very simple manner from the solution \eqref{solution2} as follows. First of all, we note that the Boussinesq equation is a higher order equation of the form
\begin{equation}
v_{tt} - v_{xx} - 3 \left(v^{2}\right)_{xx} - v_{xxxx} = 0.\label{boussinesq}
\end{equation}
Nonetheless, if we assume that it has a travelling solution of the form
\begin{equation}
v_{\rm\ssc T} = v (kx - \wt\omega t),
\end{equation}
then, following the steps in \eqref{integrated}-\eqref{further}, it can be brought to the form \eqref{integrated1} with
\begin{equation}
\widetilde{V} (v) = \frac{1}{2}\left(\left(\frac{\wt\omega}{k}\right)^{2} - 1\right) v^{2} - v^{3} + \widetilde{\beta} v + \widetilde{\gamma}.\label{boussinesqpotential}
\end{equation}
Furthermore, as we have already seen, under the map \eqref{map1},
\begin{equation}
\frac{V (g(v))}{\left(g' (v)\right)^{2}} = 2 k^{2}v^{2} - v^{3},
\end{equation}
which can be compared with \eqref{boussinesqpotential} and it is clear that in this case we can identify ${\widetilde V}(v)$ as in (\ref{map1a})
provided
\begin{equation}
\widetilde{\beta} = \widetilde{\gamma} = 0,\quad \wt\omega = \pm k \sqrt{1 + 4k^{2}}.
\end{equation}
The nontrivial travelling solution for the Boussinesq equation now follows from \eqref{map1} to be
\begin{equation}
v_{\rm\ssc T} = g^{-1} (u (kx - \wt\omega t)) = 2k^{2}\ {\rm sech}^{2}\ k (x\mp \sqrt{1+4k^{2}}\ t).\label{boussinesqsolution}
\end{equation}
This is indeed the single soliton solution of the Boussinesq equation and we note that unlike the soliton solution for the KdV (or the mKdV) equation, that of the Boussinesq equation is bi-directional.

\section{Soliton solutions from solitons}

In the last section, we tried to illustrate the details of the method of our proposal by constructing nontrivial soliton solutions of nonlinear integrable systems from a simple trigonometric solution of a linear equation. In this section, we continue the exposition of the method by showing how one can construct soliton solutions for other integrable systems starting with a soliton solution of a given nonlinear integrable system. For simplicity, let us assume that we start with the one soliton solution \eqref{kdvsolution} (or \eqref{soliton})
\begin{equation}
u_{\rm\ssc T} = 2k^{2}\ {\rm sech}^{2}\ k (x-4k^{2}t),\label{onesoliton}
\end{equation}
of the KdV equation \eqref{kdv}
\begin{equation}
u_{t} + 6 uu_{x} + u_{xxx} = 0.
\end{equation}
The potential for the KdV equation following the procedure \eqref{integrated}-\eqref{further} has already been constructed in \eqref{kdvpotential}. However, for the soliton solution in \eqref{onesoliton}, we note that both the solution as well as its derivatives vanish asymptotically which determines the constants of integration to be trivial (which we have already noted in the last section). As a result, for this case, we have
\begin{equation}
V (u) = 2k^{2} u^2 - u^{3}.\label{integratedsoliton}
\end{equation}

Let us construct a travelling solution of the negative mKdV equation \eqref{mkdv-} starting from the solution \eqref{onesoliton}. We have already seen in \eqref{mkdv-potential} that for a travelling solution of \eqref{mkdv-}
\begin{equation}
\widetilde{V} (v) = \frac{\wt\omega}{2k}\ v^{2} + \frac{1}{2}\ v^{4} + \widetilde{\beta} v + \widetilde{\gamma}.
\end{equation}
It is clear now that under the map
\begin{equation}
u (kx - \wt\omega t) = g (v (kx - \wt\omega t)) = 2 (k^{2} - v^{2}),\label{map4}
\end{equation}
we can identify 
\begin{equation}
\frac{V (g(v))}{\left(g' (v)\right)^{2}} = \frac{1}{2}\ k^{4} - k^{2} v^{2} + \frac{1}{2}\ v^{4} = \widetilde{V} (v),
\end{equation}
provided we have
\begin{equation}
\widetilde{\beta} = 0,\quad \widetilde{\gamma} = \frac{1}{2}\ k^{4},\quad \wt\omega = - 2k^{3}.
\end{equation}
This, in turn, leads to the soliton solution
\begin{equation}
v_{\rm\ssc T} = g^{-1} (u (kx - \wt\omega t)) = \pm k \tanh k (x+2k^{2}t),
\end{equation}
which we have derived earlier in \eqref{mkdv-solution}.

For the (positive) mKdV equation \eqref{mkdv+}, on the other hand, we have already determined (see \eqref{mkdv+potential})
\begin{equation}
\widetilde{V} (v) = \frac{\wt\omega}{2k}\ v^{2} - \frac{1}{2}\ v^{4} + \widetilde{\beta} v + \widetilde{\gamma}.
\end{equation}
We note that under the map
\begin{equation}
u (kx - \wt\omega t) = g (v (kx - \wt\omega t)) = 2 v^{2},\label{map5}
\end{equation}
we can identify
\begin{equation}
\frac{V (g(v))}{\left(g' (v)\right)^{2}} = \frac{1}{2}\ k^{2} v^{2} - \frac{1}{2}\ v^{4} = \widetilde{V} (v),
\end{equation}
provided we have
\begin{equation}
\widetilde{\beta} = \widetilde{\gamma} = 0,\quad \wt\omega = k^{3}.
\end{equation}
The inverse of the map \eqref{map5} now leads to the solution
\begin{equation}
v_{\rm\ssc T} = g^{-1} (u (kx - \wt\omega t)) = \pm k\ {\rm sech}\ k (x- k^{2} t),
\end{equation}
which we have also seen in \eqref{mkdv+solution}. 

At this point, it is worth comparing our construction with the conventional B\"{a}cklund transformation for KdV. We note that the Miura transformation
\begin{equation}
u = \beta v^{2} \pm \sqrt{-\beta} v_{x},\label{miura}
\end{equation}
maps the KdV equation \eqref{kdv} to the positive as well as the negative mKdV equations for $\beta=1$ and $\beta=-1$ respectively. For $\beta=1$, the transformation is clearly complex and it is worth noting that the Miura transformation alone cannot determine the time dependence of the solution (dispersion). For that one has to combine the Miura transformation \eqref{miura} with the evolution equation for mKdV to obtain the B\"{a}cklund transformation for KdV, which has the form
\begin{eqnarray}
v_{x} &=& \pm \frac{1}{\sqrt{-\beta}}\left(u - \beta v^{2}\right),\nonumber\\
v_{t} &=& \mp \frac{1}{\sqrt{-\beta}}\left(u_{xx} \pm 2\sqrt{-\beta} u_{x}v + 2u^{2} - 2\beta uv^{2}\right),\quad \beta=\pm 1.\label{backlund}
\end{eqnarray}
The compatibility of the two equations leads to the KdV equation and mapping solutions between the two systems  corresponds to working with the above set of equations which is rather nontrivial compared to our proposal. Furthermore, it seems reasonable that B\"{a}cklund transformations will map corresponding classes of solutions of the two systems into one another (namely, solitons to solitons), but it is not {\it a priori} obvious whether they can map between classes of solutions (and  even if it can be, it seems rather nontrivial). Here our proposal seems to be more versatile in that it can map solutions between different classes. In this section, we have shown how soliton solutions of KdV can generate soliton solutions of mKdV. In the next section, we will indicate how soliton solutions of KdV can also generate cnoidal solutions of mKdV. 

Let us conclude this section by indicating how the soliton solution for the Boussinesq equation \eqref{boussinesq} is obtained from \eqref{onesoliton}. We have already seen in \eqref{boussinesqpotential} that a travelling solution of the Boussinesq equation leads to
\begin{equation}
\widetilde{V} (v) = \frac{1}{2}\left(\left(\frac{\wt\omega}{k}\right)^{2} - 1\right) v^2 - v^{3} + \widetilde{\beta} v + \widetilde{\gamma}.
\end{equation}
It now follows that under a map
\begin{equation}
u (kx - \wt\omega t) = g (v (kx - \wt\omega t)) = v,
\end{equation}
we can identify
\begin{equation}
\frac{V (g(v))}{\left(g' (v)\right)^{2}} = 2 k^{2} v^{2} - v^{3} = \widetilde{V} (v),
\end{equation}
provided we have
\begin{equation}
\widetilde{\beta} = \widetilde{\gamma} = 0,\quad \wt\omega = \pm k \sqrt{1+4k^{2}}.
\end{equation}
In this case, the inverse map leads to the one soliton solution of the Boussinesq equation, namely,
\begin{equation}
v_{\rm\ssc T} = g^{-1} (u (kx - \wt\omega t)) = 2k^{2}\ {\rm sech}^{2}\ k (x \mp \sqrt{1 + 4k^{2}}\ t),
\end{equation}
which is what we have derived earlier also.

\section{Cnoidal solutions} 

Nonlinear integrable equations have a rich variety of nontrivial travelling wave solutions. In the last section we discussed one of these types of solutions, namely, the soliton solutions. There are also the cnoidal solutions which are more general than the soliton solutions and reduce to the soliton solutions in particular limits. Thus, for example, for the KdV equation \eqref{kdv}
\begin{equation}
u_{t} + 6 u u_{x} + u_{xxx} = 0,
\end{equation}
it is known that a cnoidal solution has the form
\begin{equation}
u_{\rm\ssc T} = 2m k^{2}\ cn^{2} (kx - \omega t, m) + B,\quad \omega = k (6B + 4 (2m-1)k^{2}),\label{cnoidal}
\end{equation}
where $B$ is an arbitrary constant and $m$ is a constant lying between $0\leq m\leq 1$. The $cn$ function is an elliptic function. There are other such elliptic functions which are also solutions of the KdV equation and they satisfy interesting properties. In particular, when $m=1$, the $cn$ function reduces to the ``sech" function and if we choose $m=1, B=0$, then the cnoidal solution of KdV in \eqref{cnoidal} reduces to the one soliton solution \eqref{soliton}.

Unlike the soliton solutions of the KdV equation, neither the cnoidal solutions nor their derivatives vanish asymptotically. Therefore, for such a travelling solution, following \eqref{integrated}-\eqref{further}, we can determine (see also \eqref{kdvpotential})
\begin{equation}
V (u) = \frac{\omega}{2k}\ u^{2} - u^{3} + \beta u + \gamma,\label{cnoidalpotential}
\end{equation}
where $\omega$ is defined in \eqref{cnoidal} and $\beta,\gamma$ are constants of integration whose values can be determined from the properties of the cnoidal solutions. In particular, it is known that the ``potential" in \eqref{cnoidalpotential} can be factored in terms of the three roots of the equation as
\begin{equation}
V (u) = \left(B + 2(m-1)k^{2} - u\right)\left(B-u\right)\left(B + 2mk^{2} - u\right).\label{factorization}
\end{equation}

Given the cnoidal solution of the KdV equation in \eqref{cnoidal}, let us now construct the cnoidal solution for the mKdV  equation \eqref{mkdv+} following our procedure. First, we recall from \eqref{mkdv+potential} that for a travelling solution in this case, we can write
\begin{equation}
\widetilde{V} (v)=\frac{\wt\omega}{2k}\ v^{2}-\frac{1}{2}\ v^{4} + \widetilde{\beta} v + \widetilde{\gamma}.\label{potc}
\end{equation}
Furthermore, under a map of the form
\begin{equation}
u (kx - \wt\omega t) = g (v (kx - \wt\omega t)) = 2 v^{2} + B,\label{map7}
\end{equation}
we can identify
\begin{equation}
\frac{V (g(v))}{\left(g' (v)\right)^{2}} = - \frac{1}{2}\  m (m-1)k^{4} + \frac{1}{2}\ (2m-1) k^{2} v^{2} - \frac{1}{2}\ v^{4} = \widetilde{V} (v),\label{identify1}
\end{equation}
provided we have
\begin{equation}
\widetilde{\beta} = 0,\quad \widetilde{\gamma} = - \frac{1}{2}\  m (m-1)k^{4},\quad \wt\omega = (2m-1)k^{3}.\label{const}
\end{equation}
The cnoidal solution for the mKdV equation can now be obtained from the inverse of the map \eqref{map7}
\begin{equation}
v_{\rm\ssc T} = g^{-1} (u (kx - \wt\omega t)) = \pm \sqrt{m}\ k\ cn (kx - \wt\omega t, m).\label{mkdvcnoidal}
\end{equation}
where $\wt\omega$ is given by \eqref{const}.

For $m=1$, these reduce to the soliton solutions of the (positive) mKdV equation in \eqref{mkdv+solution}. Note also that the constant introduced in the map \eqref{map7} corresponds to one of the roots of the ``potential" in \eqref{factorization}. We could instead have defined the map with any other root of the potential to obtain a different cnoidal solution of the mKdV equation. In particular, the other roots in \eqref{factorization} will  lead to solutions which will reduce to the soliton solution of the (negative) mKdV in \eqref{mkdv-solution} for $m=1.$ As we have noted in the previous section, our method can lead easily from one class of solutions to another. So, for example, starting from the one soliton solution of KdV (see \eqref{onesoliton} and \eqref{integratedsoliton}), we note that the map
\begin{equation}
u (kx -\wt\omega t) = g (v (kx -\wt\omega t)) = 2k^{2} {\rm sech}^{2}\left[cn^{-1}\left(\pm\frac{v}{\sqrt{m}k},m\right)\right],
\end{equation}
leads to \eqref{identify1}. The inverse map then generates the cnoidal solution \eqref{mkdvcnoidal} starting from a soliton solution of KdV. This demonstrates the versatility of our method.

The cnoidal solution of the Boussinesq equation \eqref{boussinesq} can also be constructed following our procedure. For a travelling solution of the Boussinesq equation, we have seen in \eqref{boussinesqpotential} that we can write
\begin{equation}
\widetilde{V} (v) = \frac{1}{2}\left(\left(\frac{\wt\omega}{k}\right)^{2} - 1\right)v^{2} - v^{3} + \widetilde{\beta} v + \widetilde{\gamma}.
\end{equation}
In this case, we see that if we define the map
\begin{equation}
u (kx - \wt\omega t) = g (v (kx - \wt\omega t)) = v,\label{map8}
\end{equation}
then we can identify
\begin{equation}
\frac{V (g(v))}{\left(g' (v)\right)^{2}} = \left(B + 2(m-1)k^{2} - v\right)\left(B-v\right)\left(B + 2mk^{2} - v\right) = \widetilde{V} (v),\label{identify}
\end{equation}
provided we have
\ben
\widetilde{\beta}&=& 4m(1-m)k^4+4B(1-2m)k^2-3B^2\nonumber
\\
\widetilde{\gamma} &=& 4Bm(m-1)k^4+2B^2(2m-1)k^2+B^3,\quad \wt\omega = \pm k\sqrt{1+ 6B + 4(2m-1)k^{2}}.\label{constants1}
\een
The inverse of the map \eqref{map8} now yields the cnoidal solution of the Boussinesq equation
\begin{equation}
v_{\rm\ssc T} = g^{-1} (u (kx - \wt\omega t)) = 2mk^{2}\  cn^{2}\ (kx-\wt\omega t, m) + B,\label{cnoidalboussinesq}
\end{equation}
where $\wt\omega$ is given by \eqref{constants1}.

Finally, let us close this section by noting that just as the soliton solutions of nonlinear systems can be obtained from travelling solutions of a linear system such as \eqref{linear}, so can the cnoidal solutions be obtained from travelling solutions of a linear system. Without going into details, let us indicate here how this can be carried out in obtaining the cnoidal solution of the KdV equation \eqref{kdv} starting with the simple trigonometric solution, \eqref{solution2}, of the third order linear equation \eqref{linear}. Let us choose $\gamma ={A^2k^{2}}/{2}$ in \eqref{potential} (as before) and note that under the map
\be\label{mapc}
u(kx-\wt\omega t)=g(v(kx-\wt\omega t))=A\cos\Biggl[cn^{-1}\left(\sqrt{\frac{v-B}{2mk^2} },m\right)\Biggr], 
\ee
we can make the identification \eqref{identify} with $\widetilde{V}(v)$ given in \eqref{kdvpotential}, provided we choose the constants
$\widetilde{\beta}$ and $\widetilde{\gamma}$ as in \eqref{constants1}, and  
\begin{equation}
\wt\omega =k(6B+4(2m-1)k^2).
\end{equation} 
In this case, the inverse of the map \eqref{mapc} leads  to the cnoidal solution \eqref{cnoidal} of the KdV equation. Similarly, the cnoidal solution \eqref{mkdvcnoidal} for the mKdV equation \eqref{mkdv+} can be generated from this solution of the linear equation under the map
\begin{equation}
u (kx - \wt\omega t) = g (v (kx - \wt\omega t)) = A\cos\Biggl[cn^{-1}\left(\frac{v}{\sqrt{m}\ k },m\right)\Biggr].
\end{equation}
This demonstrates that cnoidal solutions, like the soliton solutions, can also be obtained from simple solutions of linear systems, using our method.

\section{Some new nonlinear equations and their travelling solutions}

So far we have used our method to construct (known) solutions of known equations. We now turn our attention to applying this method to solve new nonlinear equations. We first consider a dynamical system described by
\begin{equation}
v_{t} + a (1+p)(1+2p) v^{1/p}\ v_{x} + v_{xxx} = 0,\label{pkdv}
\end{equation}
where $a$ and $p$ are arbitrary, real constants. We name this the pKdV equation, and note that for $a=p=1$ it reduces to the KdV equation while for $a=\pm2, p=1/2$, it leads to the two mKdV equations. We analyze this new equation with the method discussed in section {\bf II} and construct its travelling wave solutions starting from that of the KdV equation. For a travelling solution of \eqref{pkdv}, we can determine (following \eqref{integrated}-\eqref{further})
\begin{equation}
\widetilde{V} (v) = \frac{\wt\omega}{2k}\ v^{2} - ap^{2} v^{(1+2p)/p} + \widetilde{\beta} v + \widetilde{\gamma}.
\end{equation}
We note that under the map
\begin{equation}
u (kx - \wt\omega t) = g (v(kx - \wt\omega t)) = a v^{1/p},\label{map6}
\end{equation}
with the form of $u$ given in \eqref{form4}, we can identify
\begin{equation}
\frac{V (g(v))}{\left(g' (v)\right)^{2}} = 2k^{2}p^{2} v^{2} - ap^{2} v^{(1+2p)/p} = \widetilde{V} (v),
\end{equation}
provided we have
\begin{equation}
\widetilde{\beta} = \widetilde{\gamma} = 0,\quad \wt\omega = 4k^{3}p^{2}.
\end{equation}
In this case, the inverse of the map \eqref{map6} leads to the solution
\begin{equation}
v_{\rm\ssc T} = g^{-1} (u (kx - \wt\omega t)) = \left(\frac{2k^{2}}{a}\right)^{p}\ {\rm sech}^{2p}\ k (x - 4k^{2}p^{2} t).
\end{equation}
We note that a real solution exists for any value of $p$ when $a>0$. However, if $a<0$, the values of $p$ for which real solutions exist are restricted. For the particular value $a=1,p=1$ as well as for $a=2, p=1/2$, this solution nicely reduces to \eqref{kdvsolution} and \eqref{mkdv+solution} respectively.

Let us next construct the solutions of a different family of nonlinear equations as another illustration. We start with the negative mKdV equation \eqref{mkdv-} and we consider the kink-like soliton solution
$u_{\rm\ssc T}=\pm k \tanh k (x+2k^{2}t).$ We are then led to
\be
V(u)=\frac12(u^2-k^2)^2.
\ee
We introduce the map
\be
u(kx-\widetilde{w}_{m,a}\;t)=g(v(kx-\widetilde{w}_{m,a}\;t))=\pm k\;\cos\left(a\;\arccos(v)-m\pi\right),
\ee
where $a$ is a real parameter, and $m$ takes integer or semi-integer values. For $m$ semi-integer we can identify
\be\label{vsin}
\frac{V(g(v))}{(g'(v))^2}=\frac12\frac{k^2}{a^2}(1-v^2)\;T^2_{a}(v)=\widetilde{V}_{1,a}(v),
\ee
and for $m$ integer we have
\be\label{vcos}
\frac{V(g(v))}{(g'(v))^2}=\frac12\frac{k^2}{a^2}(1-v^2)\;U^2_{a}(v)=\widetilde{V}_{2,a}(v),
\ee
where we have denoted $T_a(v)=\cos(a\arccos(v))$ and  $U_a(v)=\sin(a\arccos(v)),$ for simplicity. For $a$ integer, $T_a(v)$ and $U_{1+a}(v)$ are the Chebyshev polynomials of the first and the second kinds, respectively. We apply our procedure to the two families of equations
\be
v_t+ (f_{i,a}(v))'\;v_x+v_{xxx}=0,
\ee
where $f_{i,a}(v)$ is a function of $v,$ and $i=1$ or $2$ correspond to the two possibilities in \eqref{vsin} and \eqref{vcos} respectively. We name this the KdV($i,a$) family, and note that for travelling waves we have
\be\label{vxxCheb}
\frac{\widetilde{w}_{i,a}}{k}\;v-f_{i,a}(v)+\wt\beta_{i,a}=(\widetilde{V}_{i,a}(v))',
\ee
with solutions
\be
v_{\rm\ssc T}=g^{-1}(u(kx-\widetilde{w}_{i,a}\;t))=\cos\left(\frac1a\left({\theta(kx-\widetilde{w}_{i,a}\;t)+m\pi}\right)\right),
\ee
where the velocitiy $\wt\omega_{i,a}$ depends on $m$ being semi-integer $(i=1)$ or integer $(i=2).$ Here $\theta(y)$ is the principal determination of $\arccos\left(\tanh(y)\right),$ and $m$ now identifies one among the several solutions the equation may comprise: for $m$ semi-integer we have $m=1/2,....,2a-1/2,$ and for $m$ integer, $m=0,1,...,2a-1,$ for $a$ being integer or semi-integer. In the figures below one identifies these possibilities very cleary. To make these illustration more explicit, let us now examine in detail the different families that we have for $a$ integer or semi-integer. 

\subsection{KdV family for $a$ semi-integer}

In order to obtain the explicit form of $(f_{i,a}(v))'$ and the related $\widetilde{w}_{i,a}$, let us write $\widetilde{V}_{i,a}(v)$ as a function of its zeroes.
For $a$ semi-integer we have
\be\label{vsemi}
\widetilde{V}_{1,a}(v)=\frac{2^{2a-3}\;k^2}{a^2}(1-v)(1-v^2)\prod^{a-\frac12}_{j=1}(v+z^j_{1,a})^2,
\ee
and $\widetilde{V}_{2,a}(v)=\widetilde{V}_{1,a}(-v)$. In the above expression we have set
\be
z^j_{1,a}=\cos\left(\frac{2j-1}{2a}\pi\right).
\ee
Using \eqref{vxxCheb} and \eqref{vsemi} we obtain
\be
(f_{1,a}(v))'=-\frac{k^2}{2a^2}-(\widetilde{V}_{1,a}(v))'',
\ee
where we have used $\widetilde{w}_{1,a}\;=\;-k^3/2a^2$. The polynomial functions \eqref{vsemi} provide a family of KdV equations which presents one  sech-like soliton solution and $N=a-1/2$ different pairs of tanh-like soliton solutions, for each member of the family of equations. We illustrate some of the many possibilities with the cases $a=1/2,3/2,$ and $5/2:$
\bes
\ben
(f_{1,1/2}(v))'=6\;k^2\;v,\qquad\qquad\widetilde{w}_{1,1/2}=-2k^3,
\een
\ben
(f_{1,3/2}(v))'=-\frac{14}{3}\;k^2\;v+\frac{80}{3}\;k^2\;v^3,\qquad\qquad\widetilde{w}_{1,3/2}=-\frac{2}{9}k^3,
\een
\ben
(f_{1,5/2}(v))'=6\;k^2\;v-\frac{114}{5}\;k^2\;v^3+\frac{672}{25}\;k^2\;v^5,\qquad\qquad\widetilde{w}_{1,5/2}=-\frac{2}{25}k^3.
\een
\ees
We also have $(f_{2,a}(v))'=-(f_{1,a}(v))'$, and $\widetilde{w}_{1,a}=\widetilde{w}_{2,a}$. In Fig.~1 we plot $\widetilde{V}_{1,a}(v)$ and the related solutions, for some semi-integer values of $a.$

\begin{figure}[!ht]
\includegraphics[{height=5cm,width=4cm,angle=-90}]{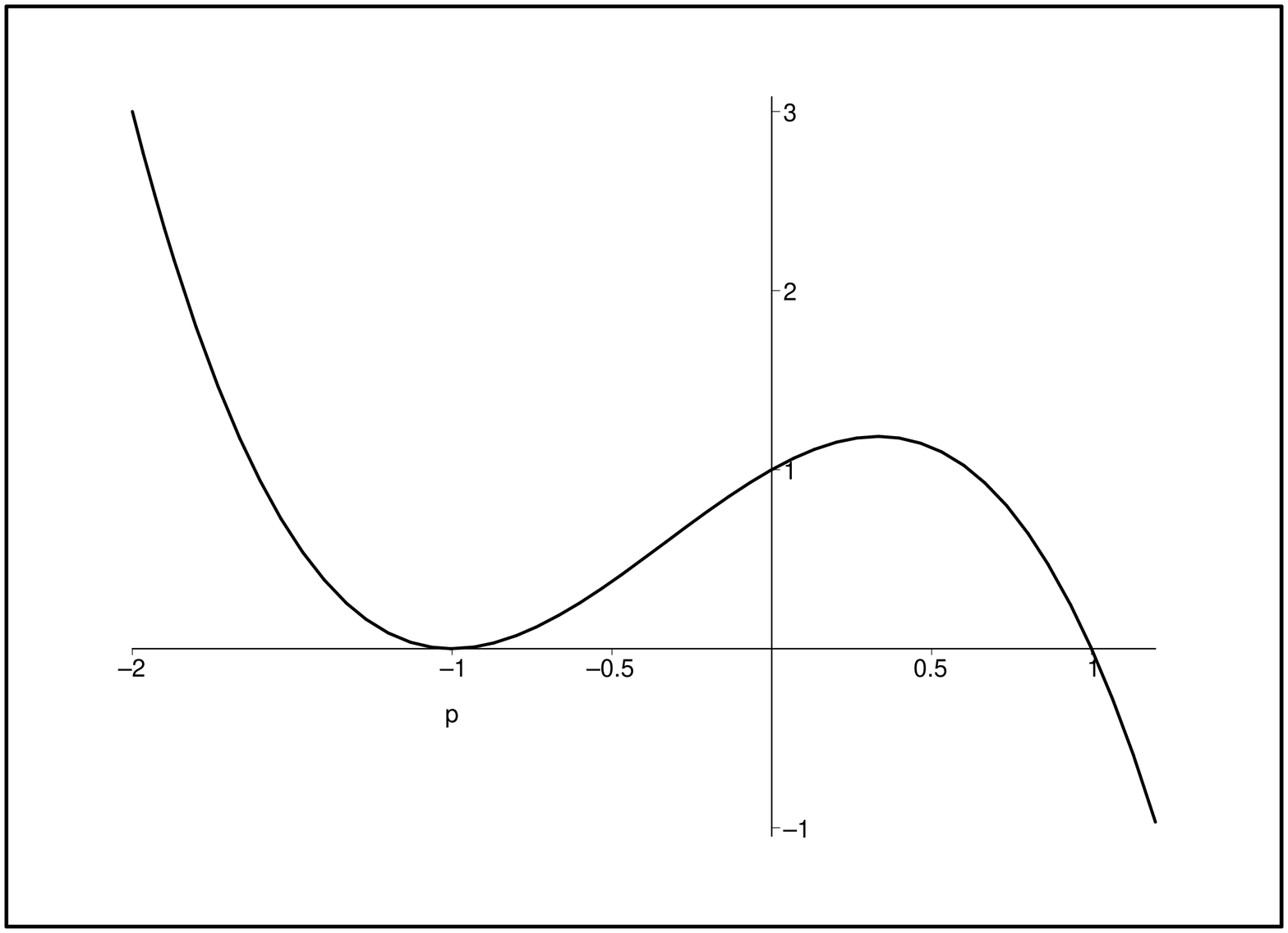}
\includegraphics[{height=5cm,width=4cm,angle=-90}]{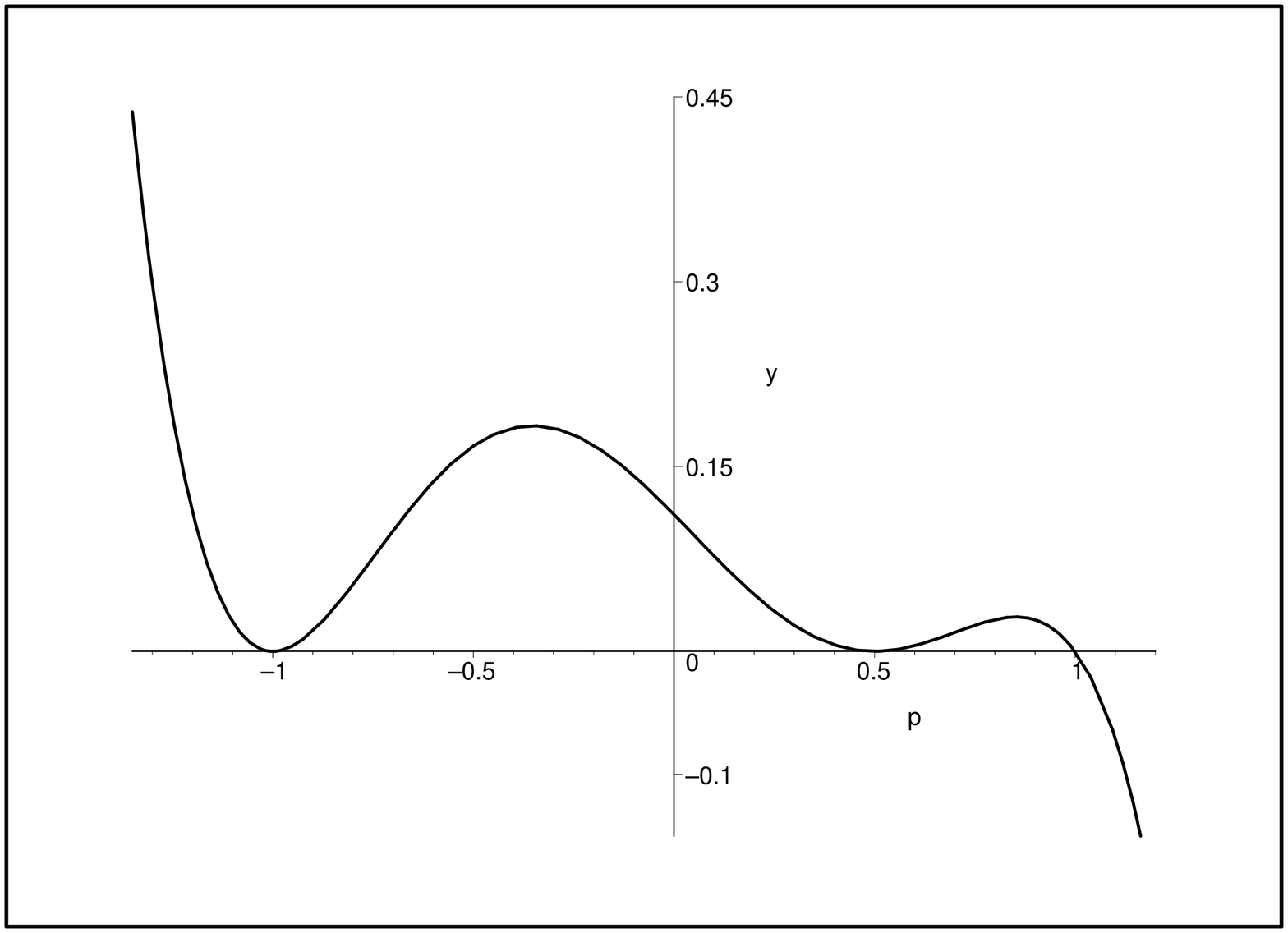}
\includegraphics[{height=5cm,width=4cm,angle=-90}]{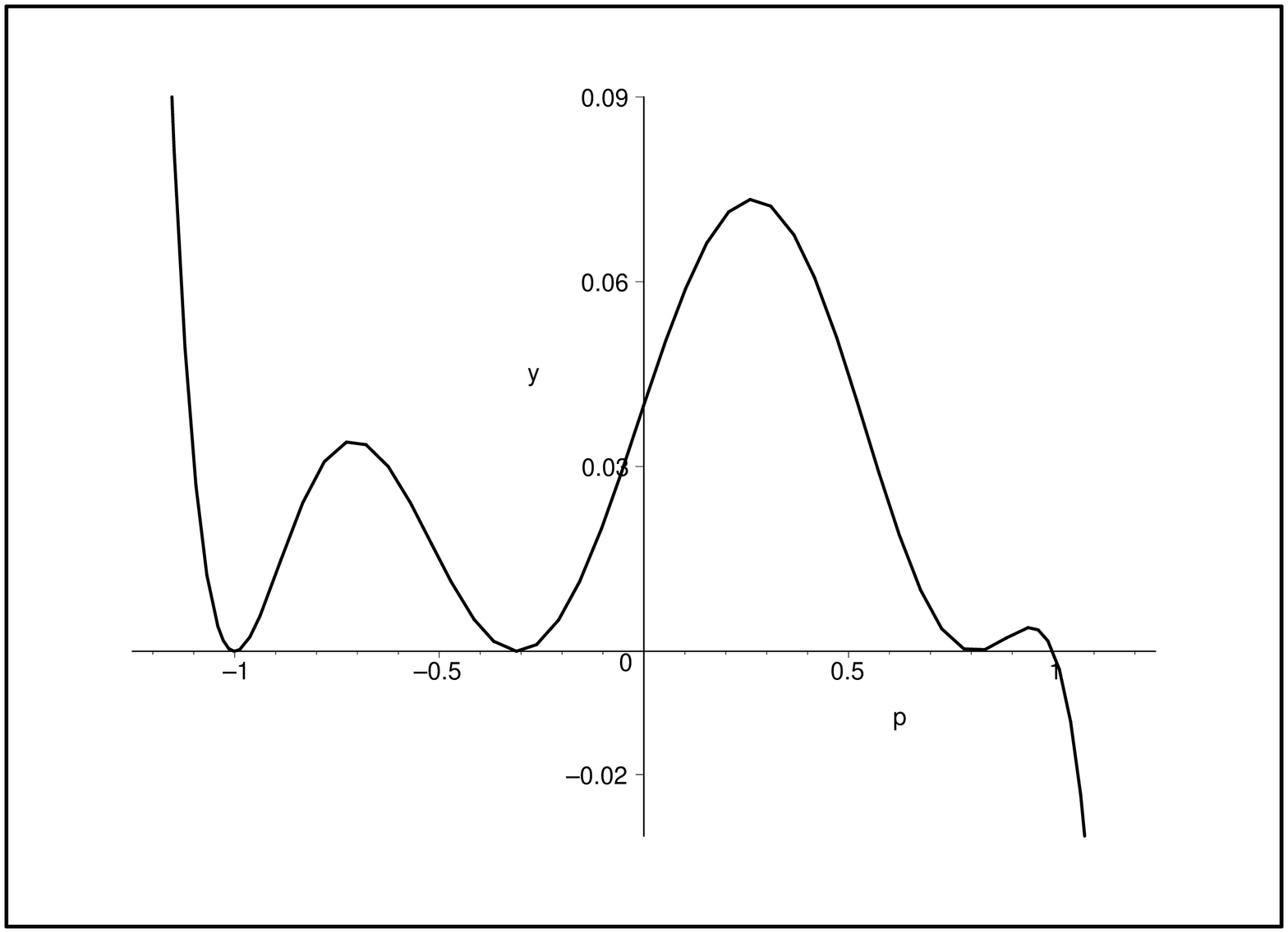}
\includegraphics[{height=5cm,width=4cm,angle=-90}]{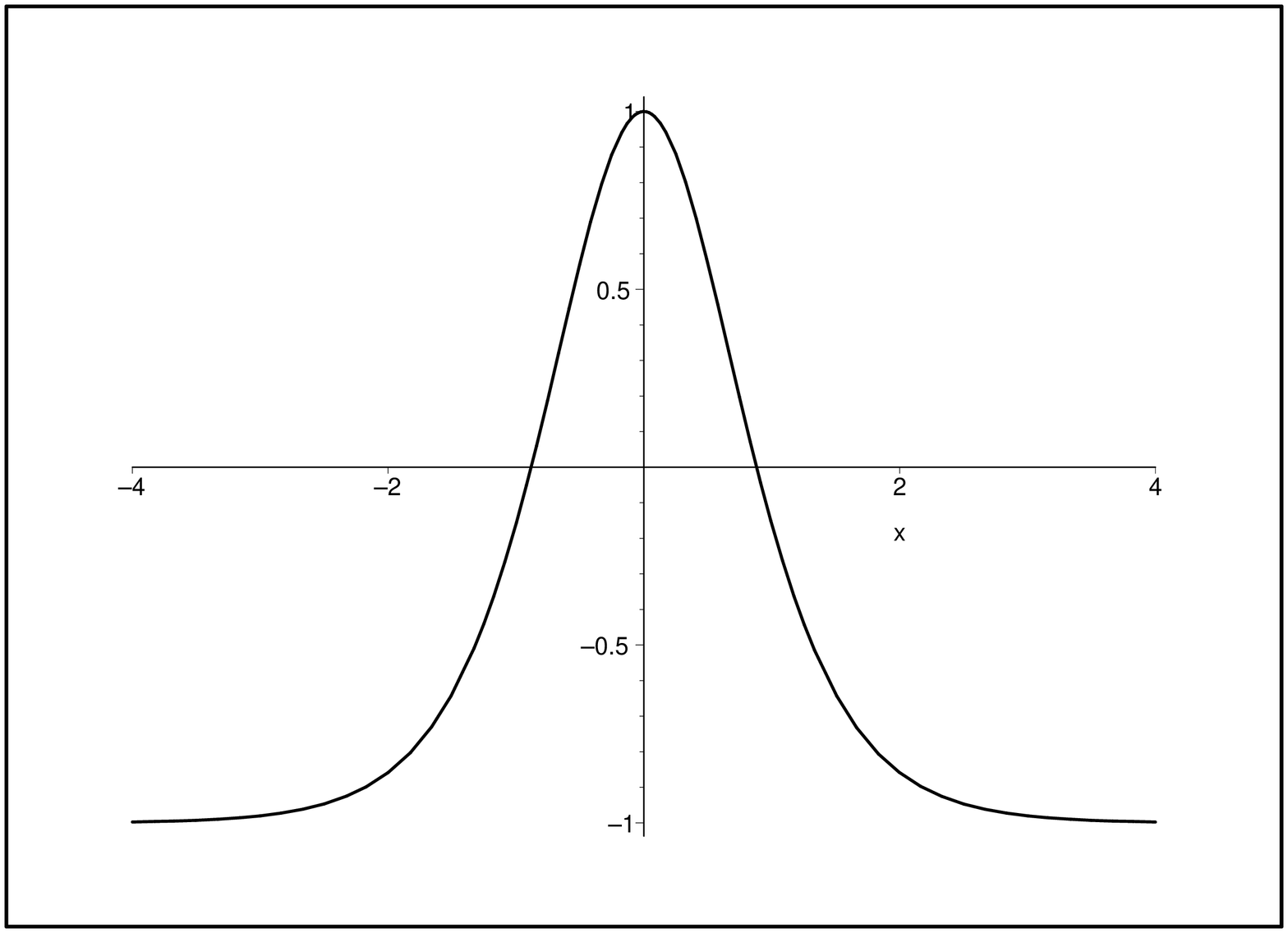}
\includegraphics[{height=5cm,width=4cm,angle=-90}]{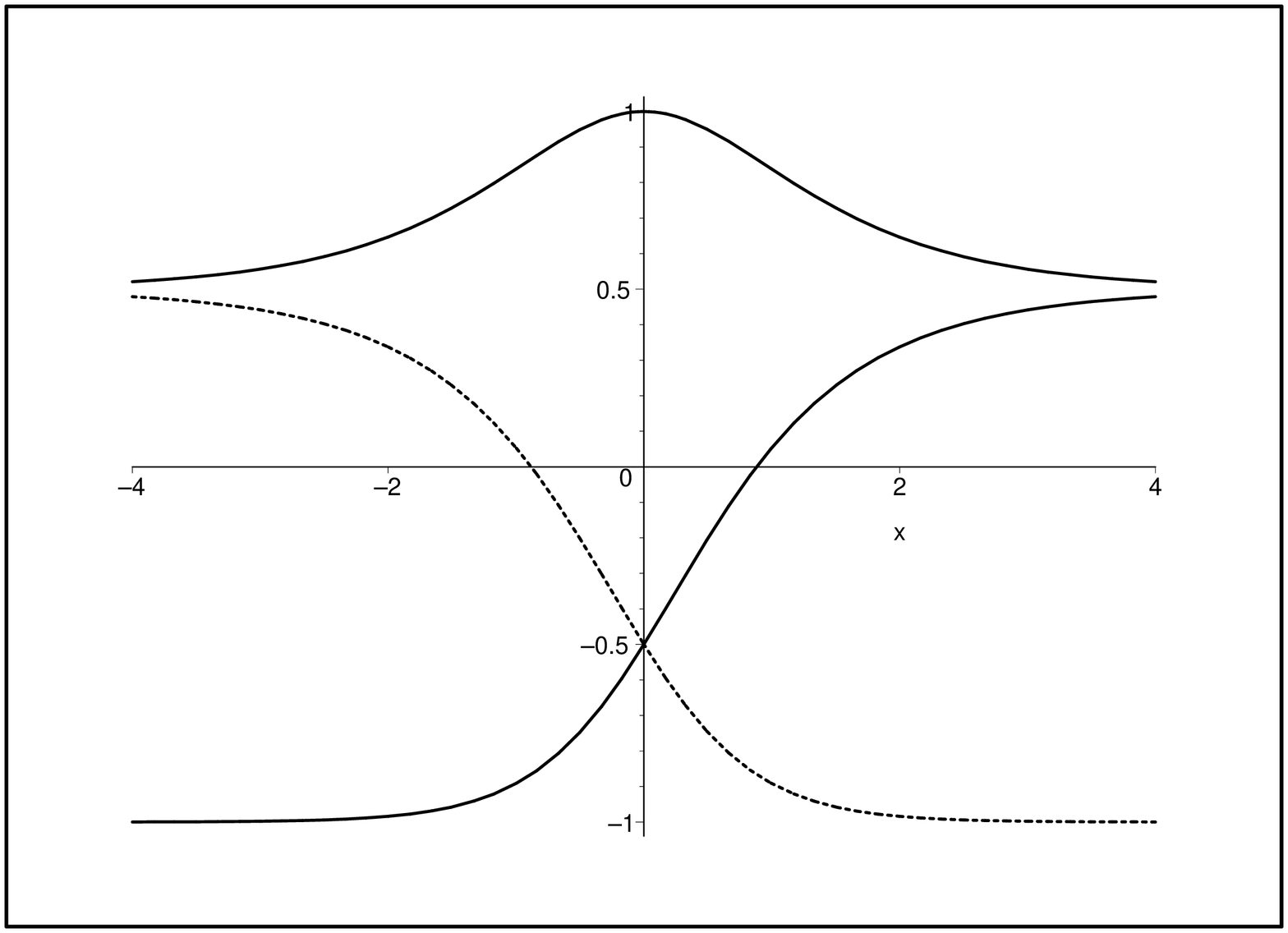}
\includegraphics[{height=5cm,width=4cm,angle=-90}]{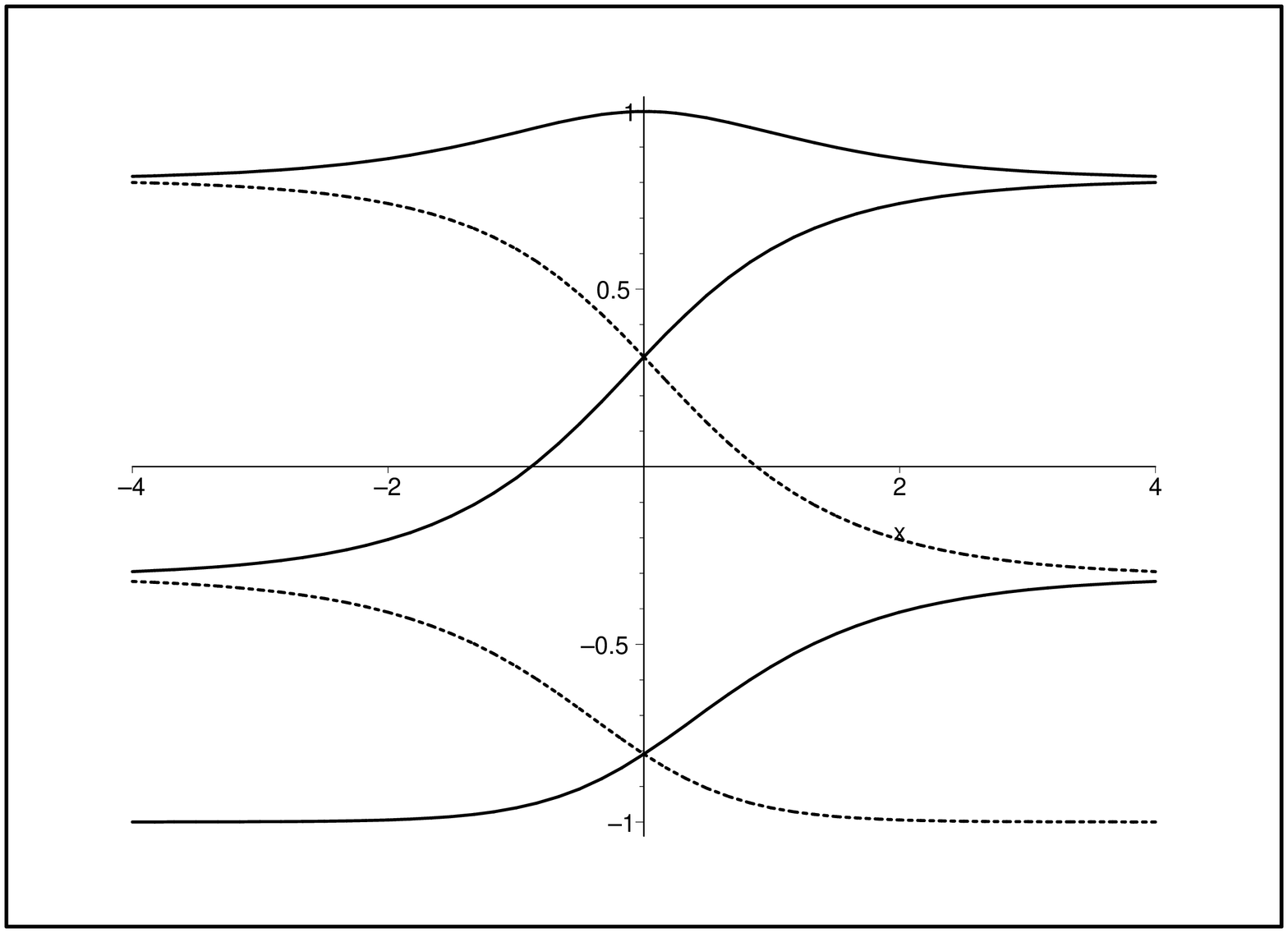}
\caption{Plots of $\widetilde{V}_{1,a}(v)$ (top panel) for $k=1$, and  $a=1/2$ (left panel), $a=3/2$ (middle panel), and $a=5/2$ (right panel), and the corresponding sech-like and tanh-like solutions (botton panel) for $t=0$.}
\end{figure}

\subsection{KdV family for $a$ integer}

In the case of $a$ integer, we have distinct families of KdV equations, one for $i=1$ and the other for $i=2.$ For this reason, below we deal with the two possibilities separately.

\subsubsection{The case $i=1$}

In order to obtain the explicit form of $(f_{1,a}(v))'$ and the related $\widetilde{w}_{1,a}$ for $a$ integer, let us write $\widetilde{V}_{1,a}(v)$ as a function of its zeroes. We have two possibilities: 

\qquad (i) $a$ even
\be\label{veven1}
\widetilde{V}_{1,a}(v)=\frac{2^{2a-3}\;k^2}{a^2}(1-v^2)\prod^{a/2}_{j=1}(v^2-(z^j_{1,a})^2)^2.
\ee

\qquad (ii) $a$ odd
\be\label{vodd1}
\widetilde{V}_{1,a}(v)=\frac{2^{2a-3}\;k^2}{a^2}v^2(1-v^2)\prod^{(a-1)/2}_{j=1}(v^2-(z^j_{1,a})^2)^2,
\ee
where
\be
z^j_{1,a}=\cos\left(\frac{2j-1}{2a}\pi\right).
\ee
Using \eqref{vxxCheb}, \eqref{veven1} and \eqref{vodd1} we obtain
\be
(f_{1,a}(v))'=\frac{\widetilde{w}_{1,a}}{k}-(\widetilde{V}_{1,a}(v))'',
\ee
with $\widetilde{w}_{1,a}=k^3$  for $a$ odd, and $\widetilde{w}_{1,a}\cong-k^3$  for $a>2$ even, with $\left|\widetilde{w}_{1,a+2}\right|<\left|\widetilde{w}_{1,a}\right|$. The polynomial functions \eqref{veven1} and \eqref{vodd1} provide families of KdV equations which present two sech-like solutions and $N=a-1$ pairs of tanh-like solutions, for each member of the family. We illustrate some of the many possibilities {\bf for} the cases $a=1,2,3:$
\bes 
\ben
(f_{1,1}(v))'=6\;k^2\;v^2,\qquad\qquad\widetilde{w}_{1,1}=k^3,
\een
\ben
(f_{1,2}(v))'=-12\;k^2\;v^2+15\;k^2\;v^4,\qquad\qquad\widetilde{w}_{1,2}=-1.25\;k^3,
\een
\ben
(f_{1,3}(v))'=22\;k^2\;v^2-\frac{200}{3}\;k^2\;v^4+\frac{448}{3}\;k^2\;v^6,\qquad\qquad\widetilde{w}_{1,3}=k^3.
\een
\ees
In Fig.~2 we plot  $\widetilde{V}_{1,a}(v)$ and the related solutions, for some integer values of $a.$
\begin{figure}[!ht]
\includegraphics[{height=5cm,width=4cm,angle=-90}]{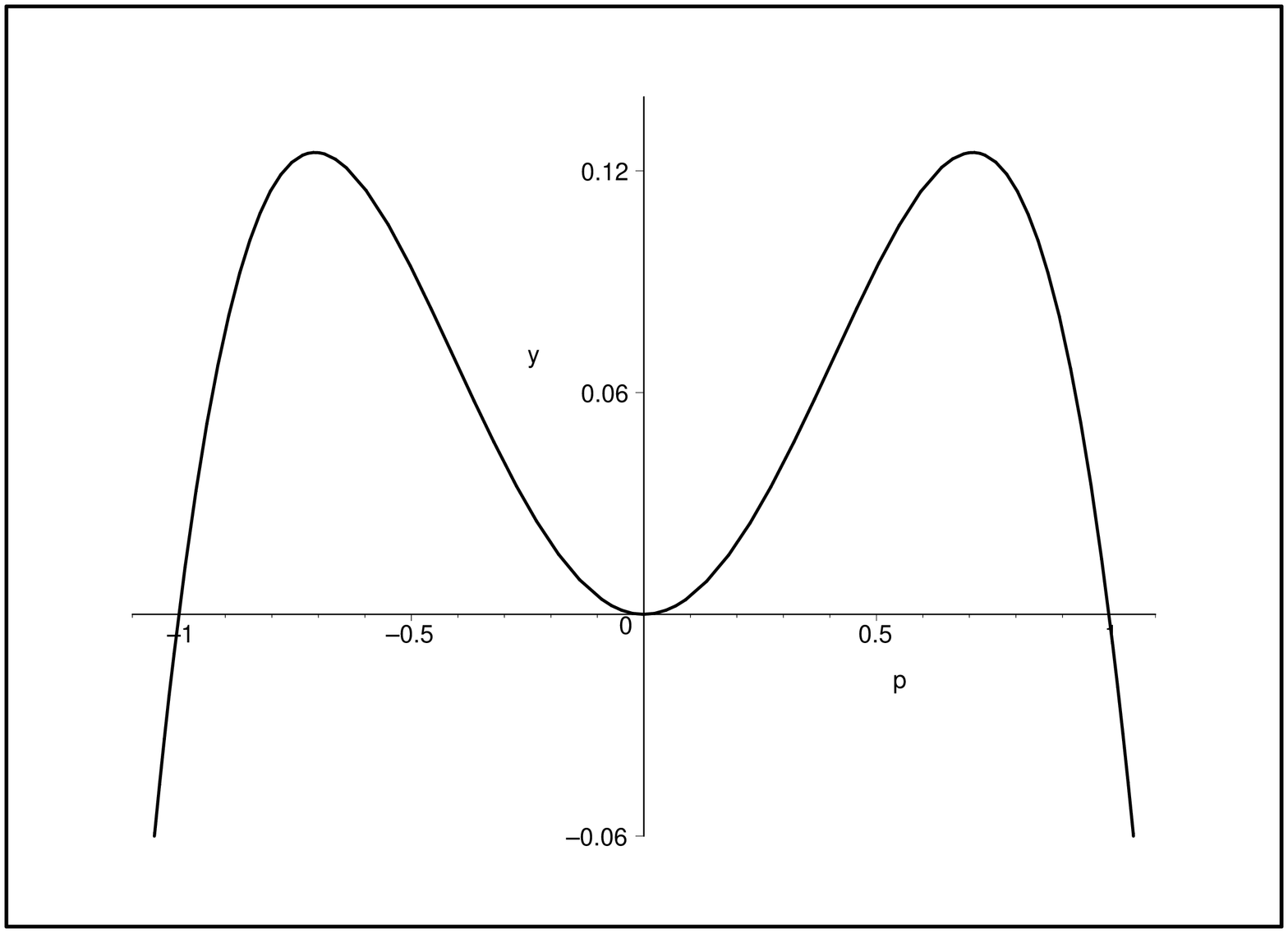}
\includegraphics[{height=5cm,width=4cm,angle=-90}]{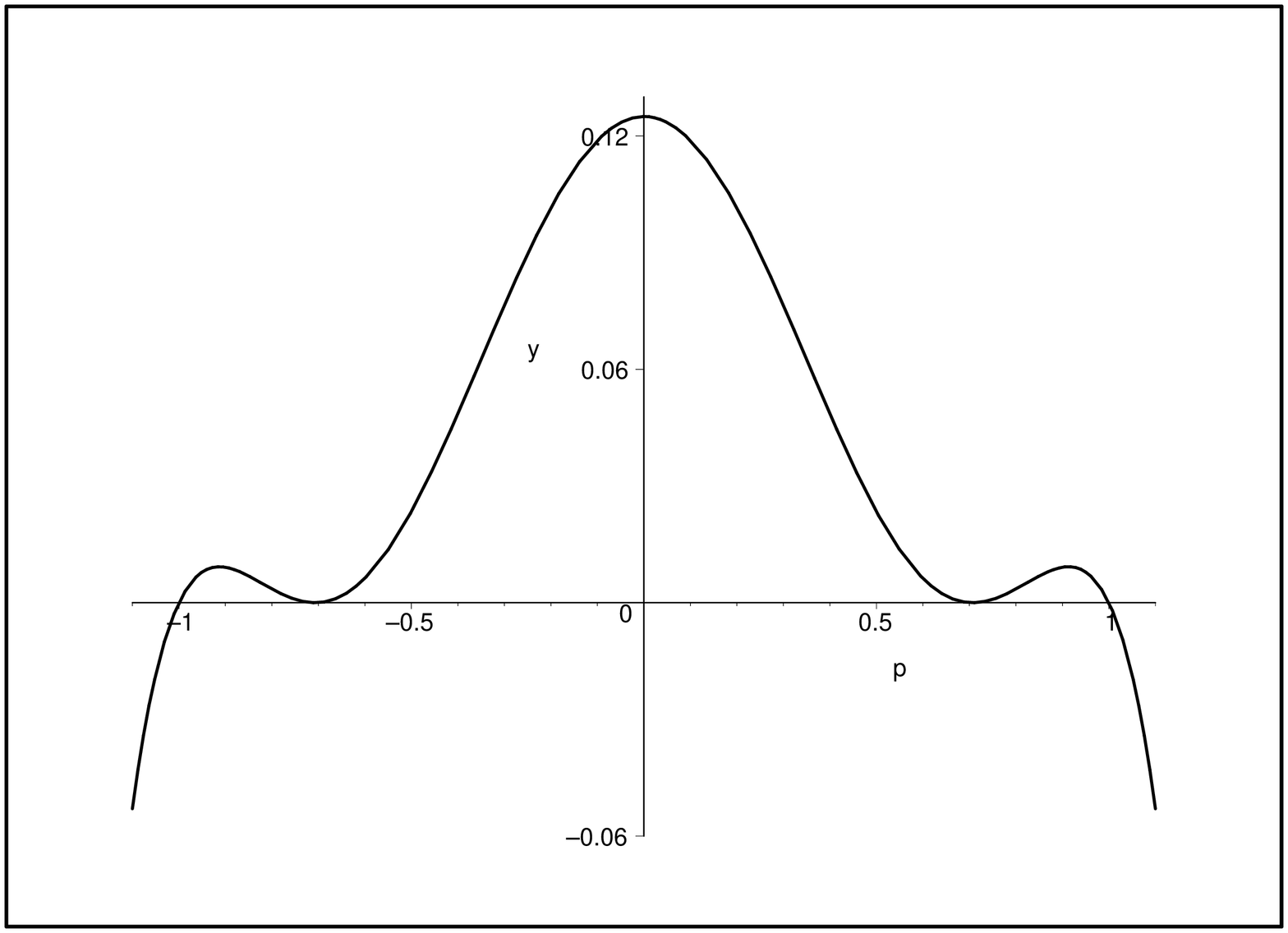}
\includegraphics[{height=5cm,width=4cm,angle=-90}]{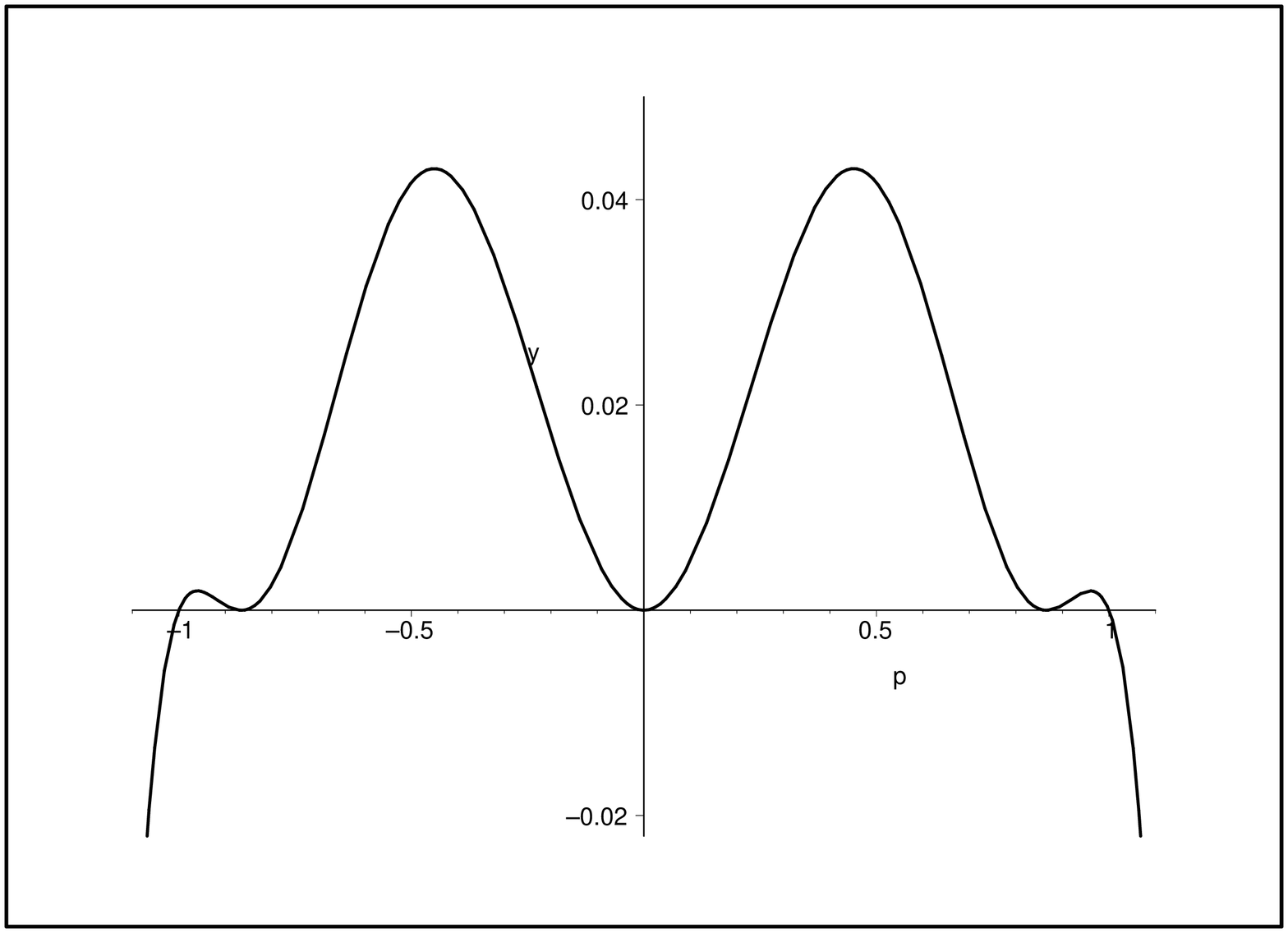}
\includegraphics[{height=5cm,width=4cm,angle=-90}]{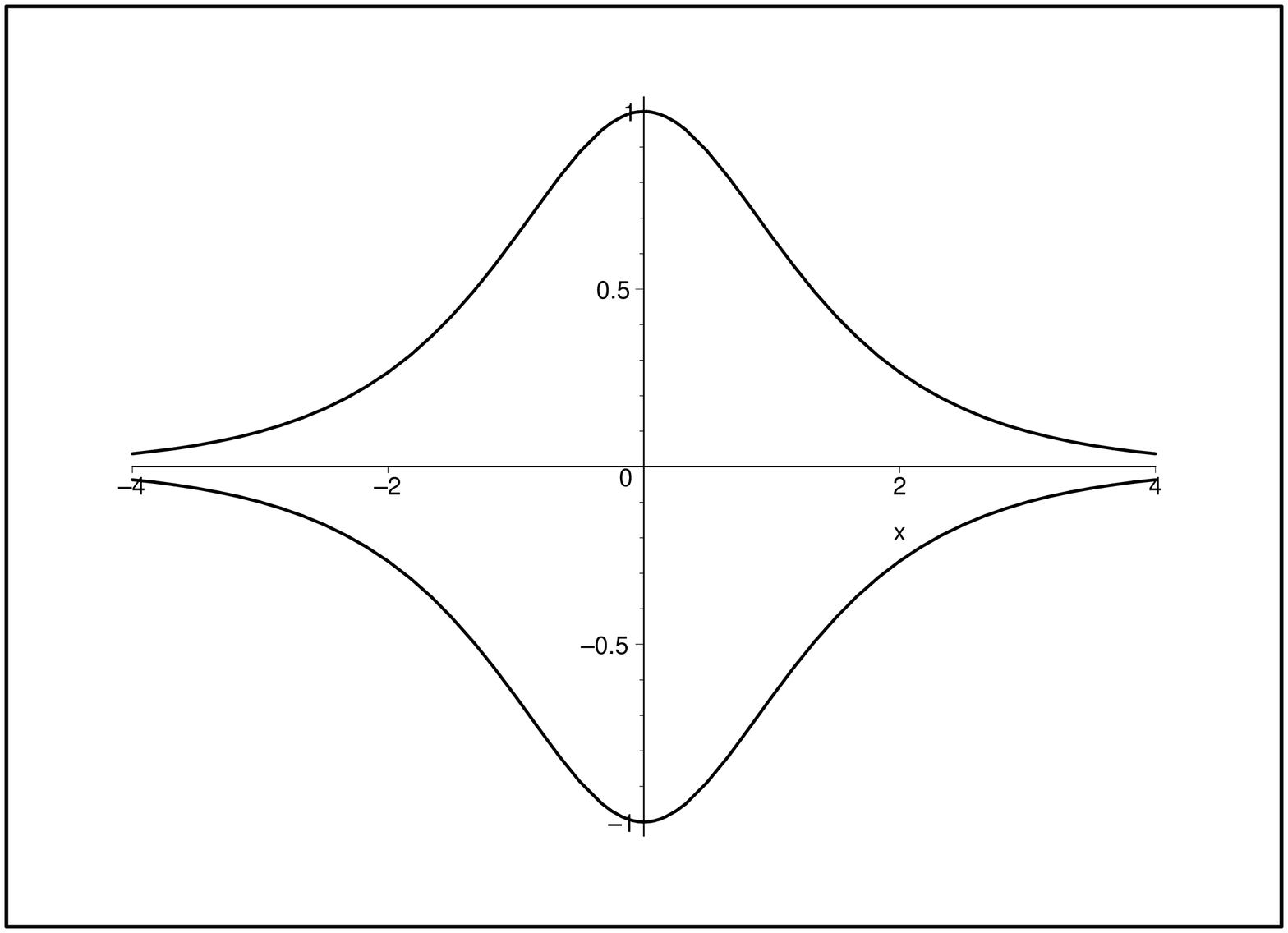}
\includegraphics[{height=5cm,width=4cm,angle=-90}]{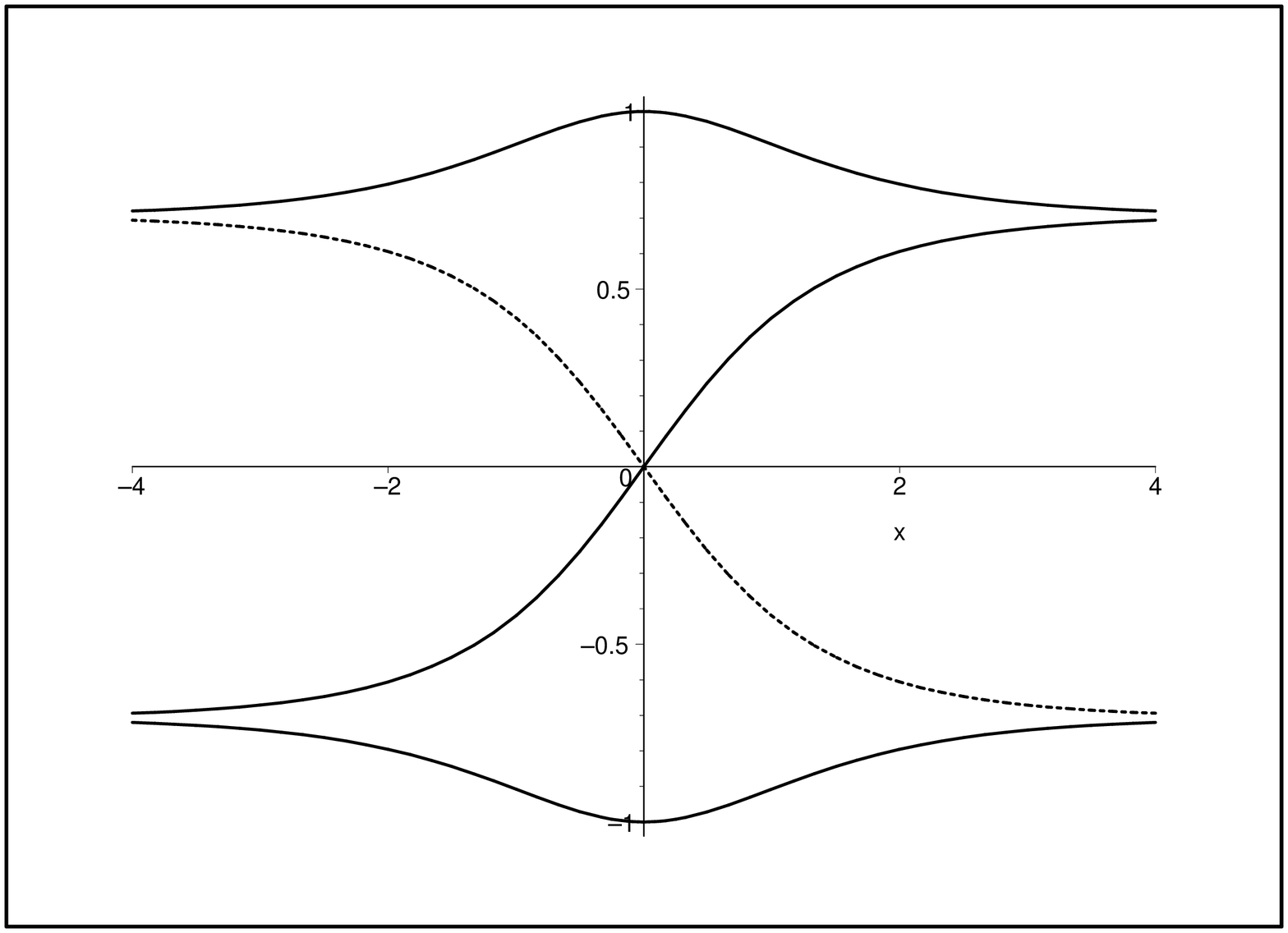}
\includegraphics[{height=5cm,width=4cm,angle=-90}]{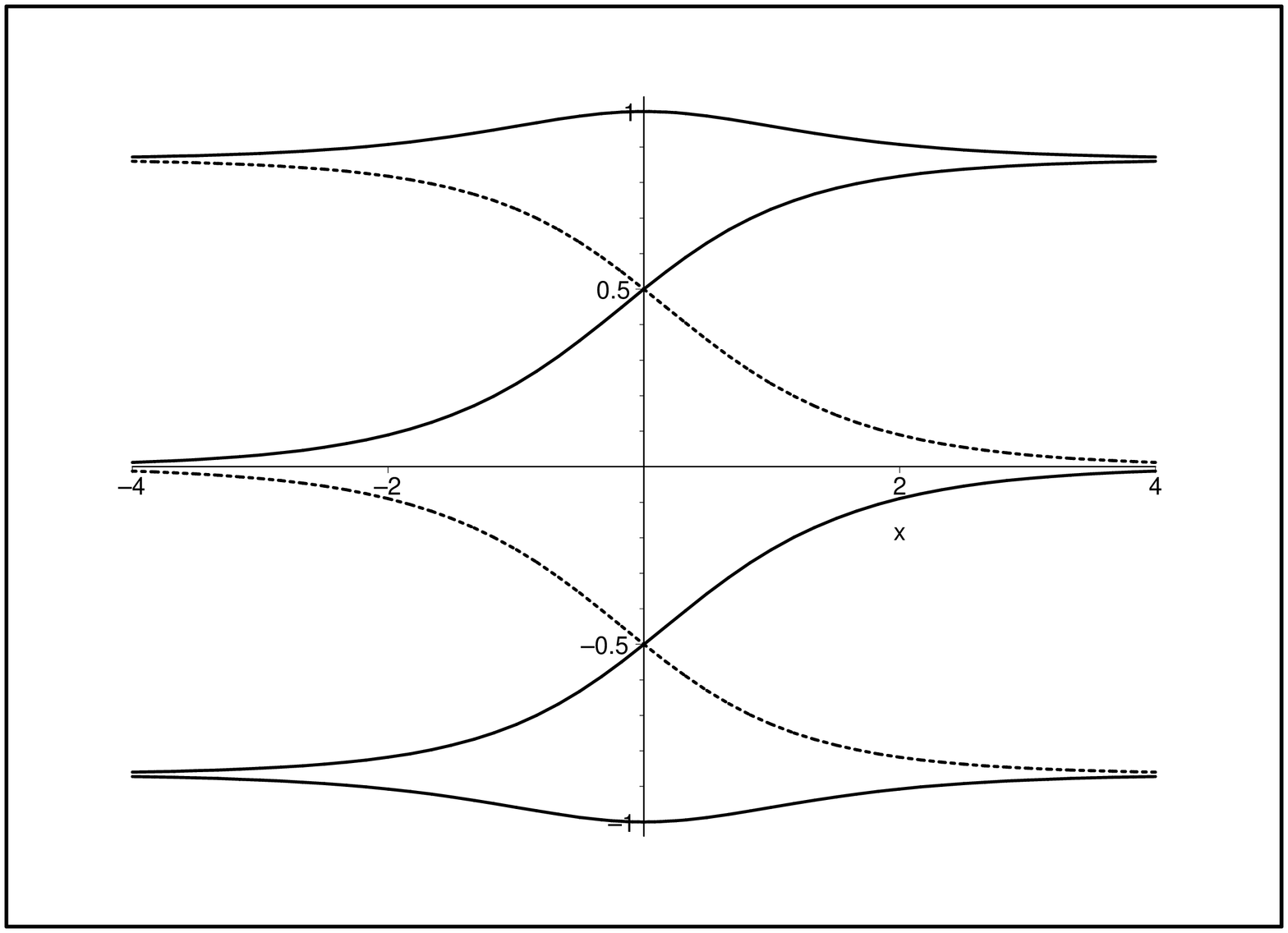}
\caption{Plots of $\widetilde{V}_{1,a}(v)$ (top panel) for $k=1$, and  $a=1$ (left), $a=2$ (middle), and $a=3$ (right),
and the corresponding solutions (botton panel) for $t=0$.}
\end{figure}

\subsubsection{The case $i=2$}

In order to obtain the explicit form of $(f_{2,a}(v))'$ and the related $\widetilde{w}_{2,a}$, let us write $\widetilde{V}_{2,a}(v)$ as a function of its zeroes.
Here we have:

\qquad (i) $a$ even
\be\label{veven2}
\widetilde{V}_{2,a}(v)=\frac{2^{2a-3}\;k^2}{a^2}v^2\prod^{a/2}_{j=1}(v^2-(z^j_{2,a})^2)^2.
\ee

\qquad (ii) $a$ odd
\be\label{vodd2}
\widetilde{V}^a_2(v)=\frac{2^{2a-3}\;k^2}{a^2}\prod^{(a+1)/2}_{j=1}(v^2-(z^j_{2,a})^2)^2,
\ee
where
\be
z^j_{2,a}=\cos\left(\frac{j-1}{a}\pi\right).
\ee
Using \eqref{vxxCheb}, \eqref{veven2} and \eqref{vodd2} we obtain
\be
(f_{2,a}(v))'=-(f_{1,a}(v))',
\ee
and $\widetilde{w}_{2,a}=k^3$  for $a$ even, $\widetilde{w}_{2,1}=-2\;k^3,$ and $\widetilde{w}_{2,a}\cong-k^3$  for $a>1,$ odd, with $\left|\widetilde{w}_{2,a+2}\right|<\left|\widetilde{w}_{2,a}\right|$. The polynomial functions \eqref{veven2} and \eqref{vodd2} provide families of KdV equations which present $N=a$ pair of tanh-like solutions, for each member of the family. In Fig.~3, we illustrate some of the many possibilities with the cases $a=1,2,3$, for $\widetilde{w}_{2,a}=-2\;k^3,k^3,-1.11\;k^3$, respectively.
\begin{figure}[!ht]
\includegraphics[{height=5cm,width=4cm,angle=-90}]{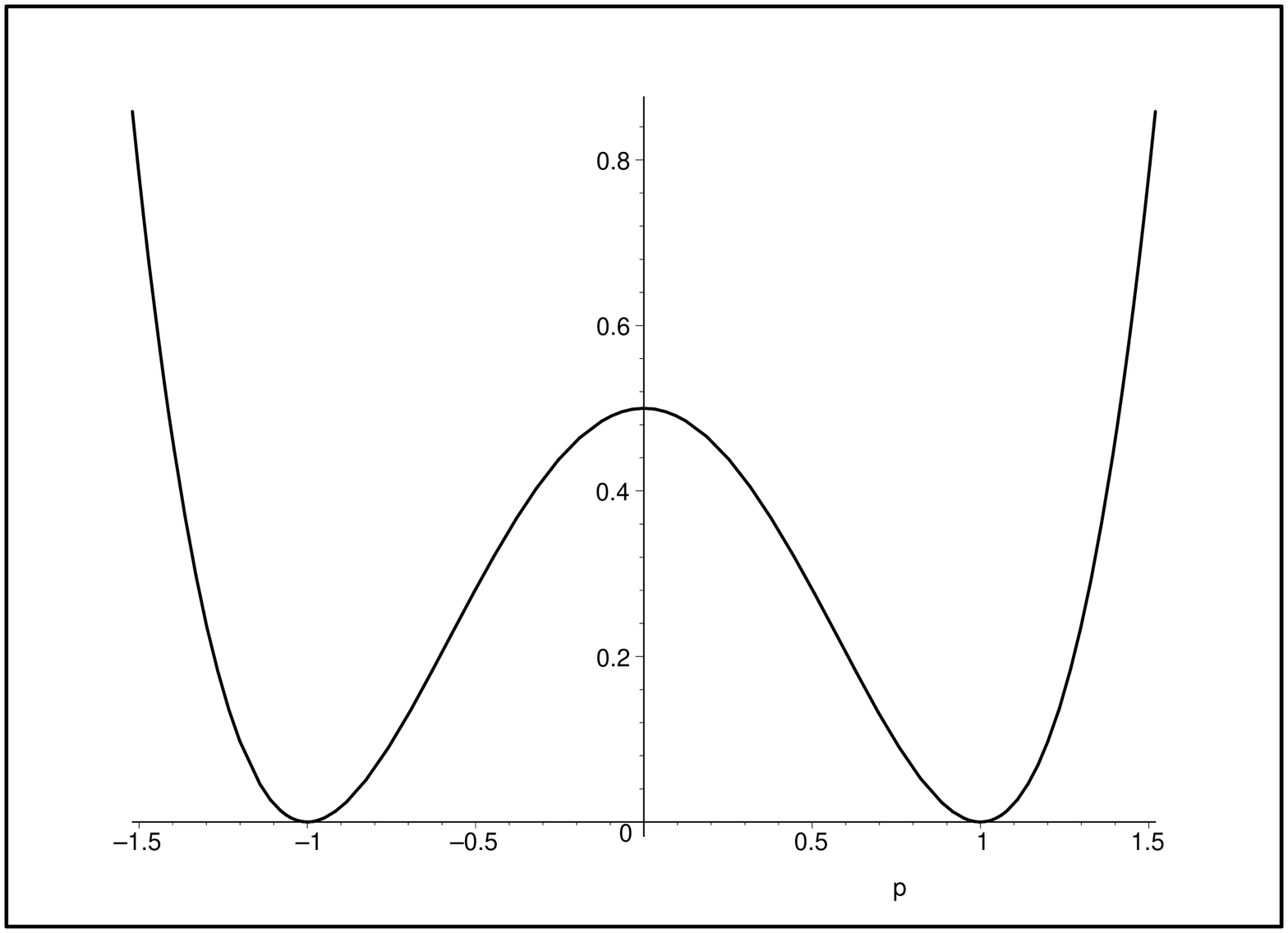}
\includegraphics[{height=5cm,width=4cm,angle=-90}]{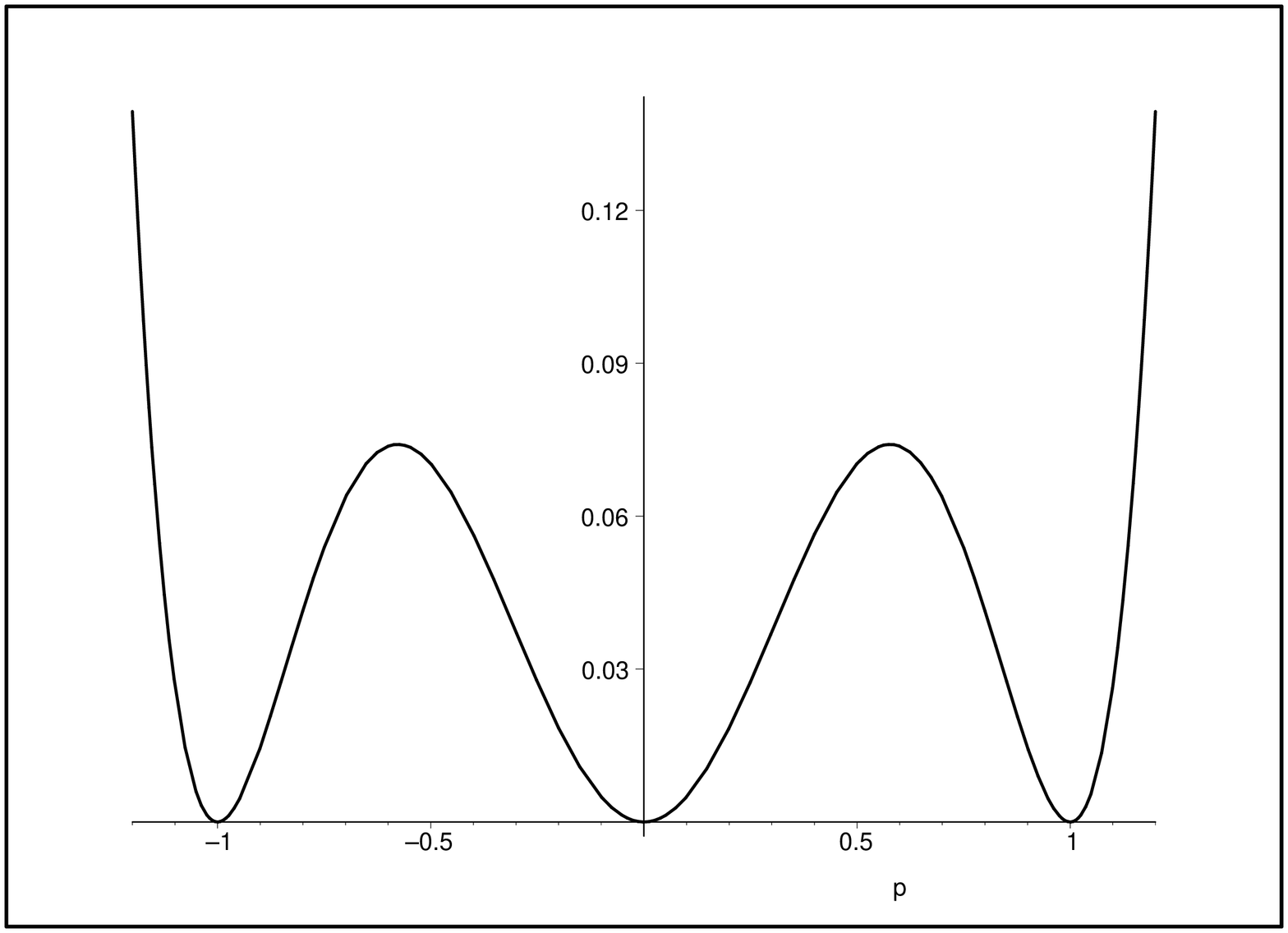}
\includegraphics[{height=5cm,width=4cm,angle=-90}]{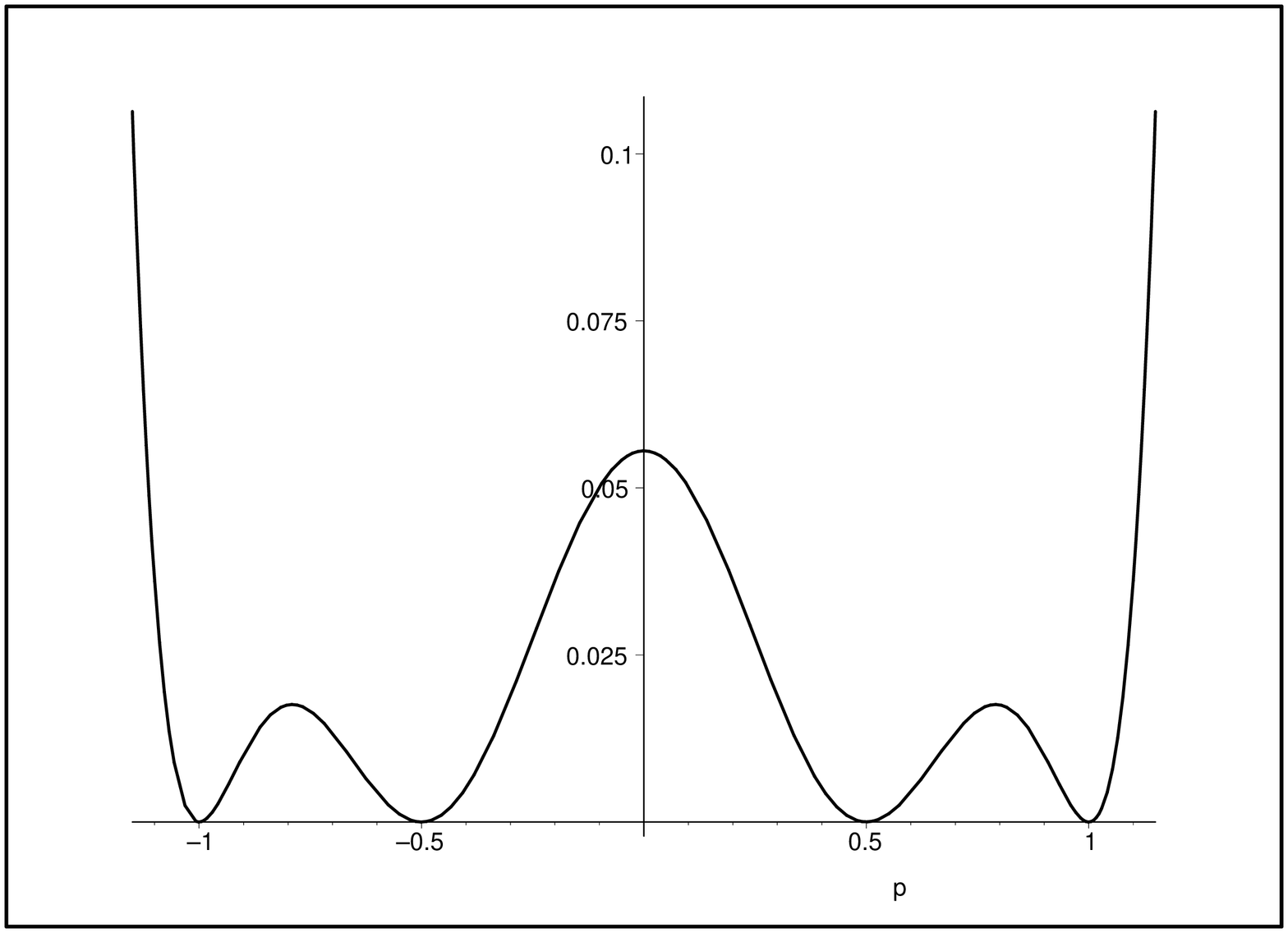}
\includegraphics[{height=5cm,width=4cm,angle=-90}]{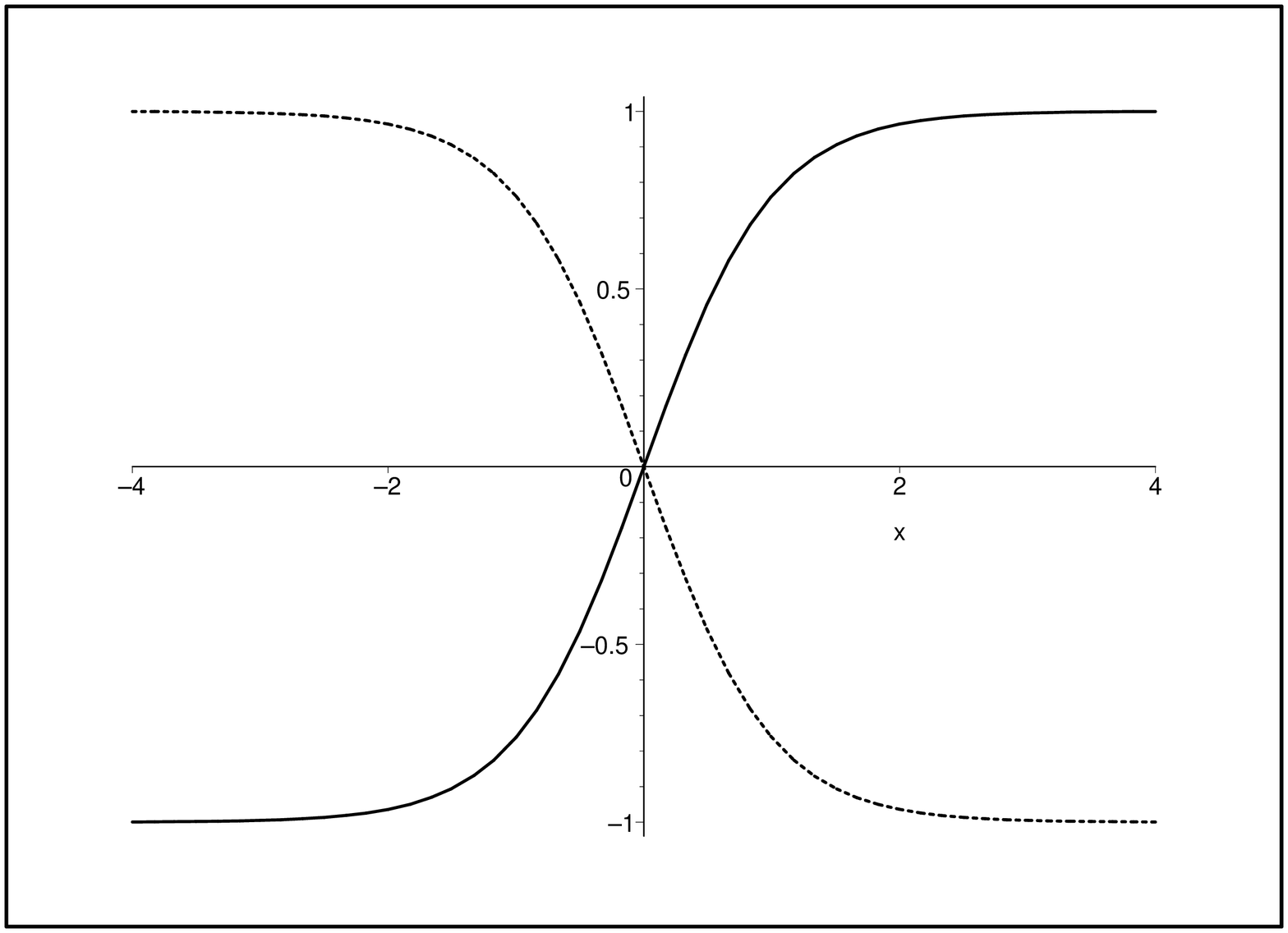}
\includegraphics[{height=5cm,width=4cm,angle=-90}]{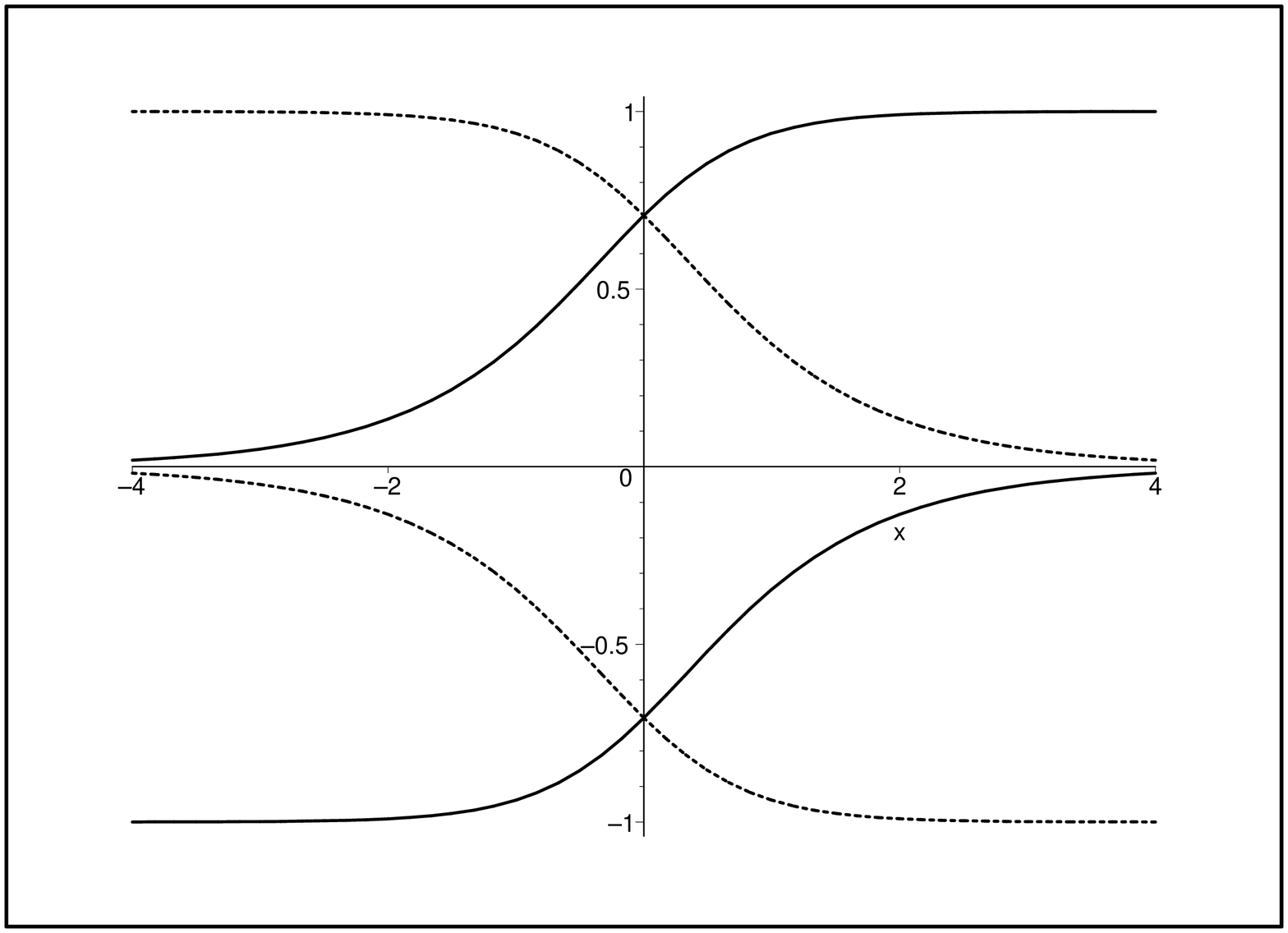}
\includegraphics[{height=5cm,width=4cm,angle=-90}]{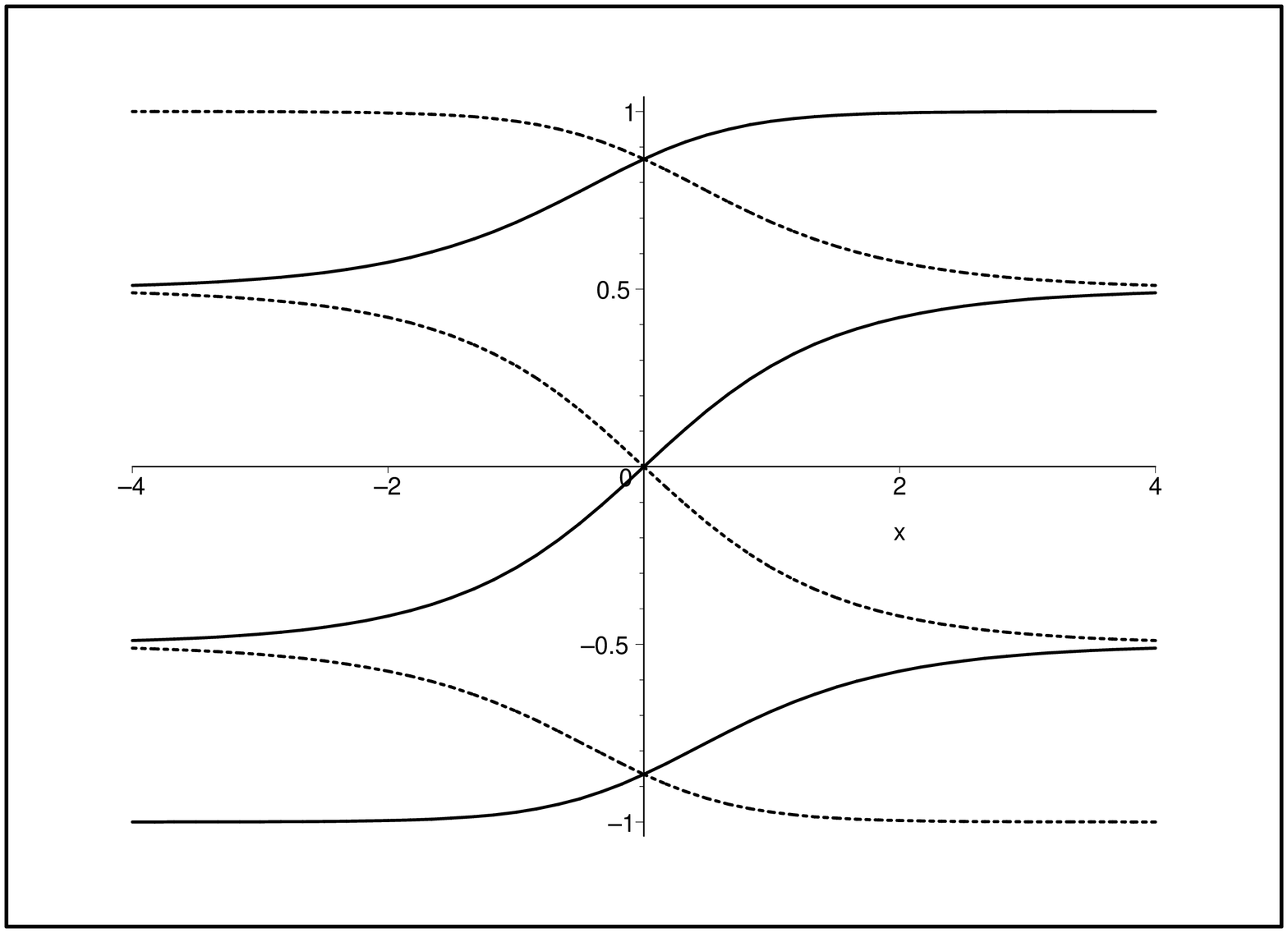}
\caption{Plots of $\widetilde{V}_{2,a}(v)$ (top panel) for $k=1$, and  $a=1$ (left panel), $a=2$ (middle panel), and $a=3$ (right panel),
and the corresponding solutions (botton panel) for $t=0$.}
\end{figure}

\section{Concluding remarks} 

In this paper, we have proposed a simple and direct method for finding travelling solutions of nonlinear integrable systems. We have illustrated the method by explicitly constructing soliton solutions of the KdV, the mKdV and the Boussinesq equations starting with a simple trigonometric solution of a linear equation. We have also shown how the method can be used profitably to construct soliton solutions from a known soliton solution of a given integrable equation. In this connection, we have constructed the soliton solutions of the mKdV and the Boussinesq equations starting from the known soliton solution of the KdV equation and have compared our method to the conventional Miura/B\"{a}cklund transformation methods. The method applies to other types of travelling solutions as well and we have emphasized this by constructing the cnoidal solutions of the mKdV and the Boussinesq equations starting with that of the KdV equation. We have also indicated how the cnoidal solutions can be obtained from simple solutions of linear systems. We have chosen these examples (of mostly known solutions) primarily to emphasize clearly how the method works in practice and how simple and useful it is. (One can contrast this with the technicalities that go into the construction of each of the known solutions.) There are many other examples that we have not been able to describe because of space limitation.

The procedure is cleary very powerful and useful. As we have explicly shown, it works very nicely to contruct and solve distinct families of given equations.
In particular, we have constructed soliton solutions for pKdV equations. We have also shown how to construct equations which support sech- and tanh-like solutions. As we know, this is in contrast with the standard situation, where the KdV equation supports sech-like soltions, and the mKdV, tanh-like solutions. The results of Sec.~{\bf VI} show that one can find sech- or tanh-like solutions for the same equation, depending on the boundary condition one uses to get the solutions. We can also check that, for the KdV hierarchy (we simply quote the result here without going into the details), the one soliton solution for the $n$-th equation in the hierarchy can be constructed using our method and has the form
\begin{equation}
u^{(2n+1)}_{\rm\ssc T} = 2k^{2}\ {\rm sech}^{2}\ k (x - (2k)^{2n} t),
\end{equation} 
where for $n=1$ (the KdV equation \eqref{kdv}) the solution corresponds to the soliton solution \eqref{soliton}. 

The results of the present work compel us to ask whether we can study multi-soliton solutions of integrabe systems, and to solve higher order equations in a systematic manner. It would be interesting, for example, if our method can shed light on whether a soliton solution exists for an equation of the form
\begin{equation}
u_{t} + \alpha uu_{x} + \delta u_{xxxxx} = 0.
\end{equation}
These are some of the questions that are presently under study and we hope to return to them in a future publication.

\bigskip

\noindent{\bf Acknowledgment}
\medskip

AD would like to acknowledge the Fulbright Foundation for a fellowship. DB and LL would like to thank CNPq for partial support. This work was supported in part by CAPES, CNPq, FAPESP, PRONEX/CNPq/FAPESQ and USDOE Grant No. DE-FG-02-91ER40685.


\end{document}